\documentclass[11pt,twoside]{article}
\usepackage[utf8]{inputenc}
\usepackage{times,fancyhdr}
\usepackage[dvips]{graphicx}
\usepackage{latexsym}
\usepackage[affil-it]{authblk}
\usepackage{amsmath}
\usepackage{amssymb}
\usepackage{mathtools}
\usepackage{color}
\usepackage[title, titletoc]{appendix}

\usepackage{hyperref}

\usepackage[backend=biber,style=phys,biblabel=brackets,pageranges=false,eprint=true]{biblatex}
\DeclareSourcemap{
 \maps[datatype=bibtex,overwrite=true]{
  \map{
    \step[fieldsource=Collaboration, final=true]
    \step[fieldset=usera, origfieldval, final=true]
  }
 }
}

\renewbibmacro*{author}{%
  \printnames{author}
  \iffieldundef{usera}{
  }{
    (\printfield{usera} Collaboration)
  }
}

\DeclareFieldFormat{eprint:arxiv}{%
    arXiv\addcolon
    \ifhyperref
    {\href{https://arxiv.org/abs/#1}{%
        #1%
    }}
    {%
        #1%
    }}

\usepackage[top=4cm, bottom=4cm, left=4cm, right=4cm]{geometry}

\begin{document}

\title{Three problems of superfluid dark matter and their solution}
\author{Tobias Mistele}
\affil{\small Frankfurt Institute for Advanced Studies\\
Ruth-Moufang-Str. 1,
D-60438 Frankfurt am Main, Germany
}
\date{}
\maketitle

\begin{abstract}
In superfluid dark matter ({\sc SFDM}), the phonon field plays a double role:
It carries the superfluid's energy density and it mediates the {\sc MOND}-like phonon force.
We show that these two roles are in tension with each other on galactic scales:
A {\sc MOND}-like phonon force is in tension with a superfluid in equilibrium and with a significant superfluid energy density.
To avoid these tensions, we propose a model where the two roles are split between two different fields.
This also allows us to solve a stability problem in a more elegant way than standard {\sc SFDM}.
We argue that the standard estimates for the size of a galaxy's superfluid core need to be revisited.
\end{abstract}

\section{Introduction}
\label{sec:introduction}

There is ample evidence for missing non-baryonic mass from observations on both cosmological and galactic scales.
The simplest explanation for the evidence from cosmological scales, most notably the Cosmic Microwave Background ({\sc CMB}), is the existence of cold dark matter particles ({\sc CDM}).
In contrast, the simplest explanation for the evidence from galactic scales, most notably rotation curves, is in terms of a modified force law of the form proposed by Modified Newtonian Dynamics ({\sc MOND}) \cite{Milgrom1983a,Milgrom1983b,Milgrom1983c,Bekenstein1984}.
The aim of superfluid dark matter ({\sc SFDM}) as proposed in Ref.~\cite{Berezhiani2015} is to combine the best of both worlds in a single model.
That is, to reproduce {\sc CDM} on cosmological scales and {\sc MOND} on galactic scales.
The idea is to postulate a new type of particle that behaves differently on different scales.
On galactic scales, the new particles condense to a superfluid, where a phonon force then exerts an additional {\sc MOND}ian force on baryons.
In contrast, these particles are in the normal, not-condensed phase on cosmological scales, where they behave like {\sc CDM} \cite{Berezhiani2015, Berezhiani2018}.
Superfluid condensates of dark matter on galactic scales have been considered before \cite{Sikivie2009,Noumi2014,Davidson2013,DeVega2014,Davidson2015,Guth2015,Aguirre2016,Dev2017,Eby2018,Sarkar2018}, but the long-range phonon force is specific to the proposal from Ref.~\cite{Berezhiani2015}.

In {\sc SFDM} as proposed in Ref.~\cite{Berezhiani2015}, the so-called phonon field plays a double role in galaxies:
It carries both the {\sc MOND}-like phonon force and the superfluid's energy density.
The phonon force is needed to reproduce {\sc MOND}ian rotation curves at small and intermediate radii, while the superfluid's energy density is needed to reproduce the strong lensing signal at larger radii \cite{Hossenfelder2019}.
In the present work, we will show that these two roles of the phonon field are in tension with each other and how these tensions can be avoided by assigning each role to a separate field.
We will also see that the two standard estimates for the size of a galaxy's superfluid core should be revisited because, in general, they do not agree with each other.
In the following, we employ units with $ c = \hbar = 1 $ and the metric signature $ (+, -, -, -) $.
Small Greek indices run from $ 0 $ to $ 3 $ and denote spacetime dimensions.

In Sec.~\ref{sec:sfdm}, we introduce the standard model of {\sc SFDM} from Ref.~\cite{Berezhiani2015} and discuss three problems of this model.
In Sec.~\ref{sec:improved}, we propose an improved model and show how this improved model avoids the problems from Sec.~\ref{sec:sfdm}.
Then, we choose explicit numerical parameters for our model in Sec.~\ref{sec:params}, which we use to illustrate the model's phenomenology on galactic scales with the Milky Way as an example in Sec.~\ref{sec:pheno}.
In Sec.~\ref{sec:RTRNFW}, we estimate the size of the Milky Way's superfluid core using standard methods and argue that these methods should be revisited.
After a short discussion in Sec.~\ref{sec:discussion} we conclude in Sec.~\ref{sec:conclusion}.

\section{Three problems of {\sc SFDM}}
\label{sec:sfdm}

Ref.~\cite{Berezhiani2015} formulates {\sc SFDM} in terms of an effective field theory ({\sc EFT}) for the phonon field $ \theta $.
This {\sc EFT} is valid on galactic scales, but may break down on cosmological or solar system scales.
Concretely, the Lagrangian reads \cite{Berezhiani2015}
\begin{subequations}
 \label{eq:L}
\begin{align}
 \mathcal{L} &= f(K_\theta - m^2) - \lambda \, \theta \, \rho_b \,, \\
 f(K_\theta-m^2) &= \frac{2 \Lambda}{3} \sqrt{|K_\theta - m^2|} (K_\theta - m^2) \,, \\
    K_\theta &= \nabla^\alpha \theta \nabla_\alpha \theta \,.
\end{align}
\end{subequations}
Here, $ m $ is the mass of the particles of which the superfluid consists, $ \Lambda $ is a constant with mass dimension 1 that is related to the self-interaction of these particles, and $ \lambda \equiv \bar{\alpha} \Lambda/M_{\rm{Pl}} $ is a dimensionless constant that couples $ \theta $ to the baryonic energy density $ \rho_b $.
We can introduce a chemical potential $ \mu $ by shifting $ \dot{\theta} \to \dot{\theta} + \mu $.\footnote{We do not shift $ \theta \to \theta + \mu \cdot t $ for reasons discussed in Ref.~\cite{Mistele2019}, see also Sec.~\ref{sec:prob:equilibrium}.}
In the non-relativistic limit, we have $ \mu = m + \mu_{\rm{nr}} $ with $ |\dot{\theta}| \ll m $ and $ \mu_{\rm{nr}} \ll m $, where $ \mu_{\rm{nr}} $ is the non-relativistic chemical potential.
In the following, it will be useful to use the notation
\begin{align}
 \hat{\mu}(\vec{x}) \equiv \mu_{\rm{nr}} - m \phi_{\rm{N}}(\vec{x}) \,,
\end{align}
where $ \phi_{\rm{N}} $ is the Newtonian gravitational potential.
Note that $ \mu_{\rm{nr}} $ is a constant, but $ \hat{\mu} $ is not due to its dependence on $ \phi_{\rm{N}} $.
For time-independent equilibrium solutions, we then have $ K_\theta - m^2 = 2 m \hat{\mu} - (\vec{\nabla} \theta)^2 $.
The acceleration of the baryons due to $ \theta $ is $ \vec{a}_\theta = - \lambda \vec{\nabla} \theta $.
As discussed in Ref.~\cite{Berezhiani2015}, this acceleration has the standard {\sc MOND} form $ |\vec{a}_\theta| \approx \sqrt{a_0 |\vec{a}_b|} $ if $ (\vec{\nabla} \theta)^2 \gg 2 m \hat{\mu} $.
Here, $ a_0 = \bar{\alpha}^3 \Lambda^2/M_{\rm{Pl}} $ is the {\sc MOND}ian acceleration scale and $ \vec{a}_b $ is the Newtonian gravitational acceleration due to the baryons.
In this case, $ K_\theta - m^2 \approx - (\vec{\nabla} \theta)^2 < 0 $.
We discuss this {\sc MOND} limit in more detail in Sec.~\ref{sec:prob:mondlimit}.

\subsection{The stability problem}
\label{sec:prob:stability}

One problem with the Lagrangian $ \mathcal{L} $ from Eq.~\eqref{eq:L} is that perturbations around the desired equilibrium solutions in galaxies are unstable.
This was already discussed in Ref.~\cite{Berezhiani2015}.
To see this, consider a perturbation $ \delta $ on top of an equilibrium background solution $ \theta_0 $ with $ K_{\theta_0} - m^2 < 0 $, e.g. a {\sc MOND} limit solution with $ 2 m \hat{\mu}_0 \ll (\vec{\nabla} \theta_0)^2 $.
Here, $K_{\theta_0}$ is the background value of $K_\theta$ with a chemical potential $\mu_0 = m + \mu_{\rm{nr}}^0$ and $\hat{\mu}_0 = \mu_{\rm{nr}}^0 - m \phi_{\rm{N}}$.
Then, the second-order Lagrangian for the perturbation $ \delta $ reads
\begin{align}
 \label{eq:Lpert}
 \mathcal{L}_{\rm{pert}} = \left[ f_0' g^{\alpha \beta} + 2 f_0'' \nabla^\alpha \theta_0 \nabla^\beta \theta_0\right] \nabla_\alpha \delta \nabla_\beta \delta \,,
\end{align}
where
\begin{align}
 \label{eq:f0}
  f_0' &= \Lambda \sqrt{|K_{\theta_0} - m^2|} > 0 \,, \\
 f_0'' &= - \frac{\Lambda}{2} \frac{1}{\sqrt{|K_{\theta_0} - m^2|}} < 0 \,.
\end{align}
For stability, the prefactor of the $ \dot{\delta}^2 $ term in $\mathcal{L}_{\rm{pert}}$ should be positive.
This prefactor is:
\begin{align}
 f_0' g^{00} + 2 \left(g^{00}\right)^2 f_0'' \mu_0^2  \approx \frac{\Lambda m^2}{\sqrt{|K_{\theta_0} - m^2|}} \left(\frac{|K_{\theta_0} - m^2|}{m^2} - 1\right) < 0 \,.
\end{align}
This term is negative because typically $ |K_{\theta_0} - m^2| \approx |2 m \hat{\mu}_0 - (\vec{\nabla} \theta_0)^2| \ll m^2 $ in the non-relativistic limit on galactic scales \cite{Berezhiani2015}.
Thus, the desired equilibrium solutions in galaxies are unstable.

To fix this instability, Ref.~\cite{Berezhiani2015} introduces finite-temperature corrections parametrized by a new parameter $ \bar{\beta} $.
In the non-relativistic limit, the Lagrangian then is:
\begin{align}
 \label{eq:Lbeta}
 \mathcal{L}_{\bar{\beta}} = \frac{2 \Lambda}{3} \sqrt{(2m) |X - \bar{\beta} Y|}  (2m X) \,,
\end{align}
where $ 2m X = 2 m (\hat{\mu} + \dot{\theta}) - (\vec{\nabla} \theta)^2 $ is the non-relativistic limit of $ K_\theta - m^2 $ and in the rest-frame of the fluid we have $ Y = \hat{\mu} + \dot{\theta} $.
This agrees with the non-relativistic limit of the original Lagrangian $\mathcal{L}$ for $\bar{\beta} = 0$.
For $ \bar{\beta} > 3/2 $, the addition of the $ \bar{\beta} Y $ term under the square root then cures the instability described above \cite{Berezhiani2015}.
However, this solution is not entirely satisfactory since both the functional form of the finite-temperature corrections and the numerical value of $ \bar{\beta} $ are chosen ad-hoc.
In principle, the finite-temperature Lagrangian $ \mathcal{L}_{\bar{\beta}} $ should follow from the zero-temperature Lagrangian $ \mathcal{L} $ from Eq.~\eqref{eq:L}, but so far this has not been established.
As a result, it is unclear how $ \bar{\beta} $ depends on the superfluid temperature and it is unclear whether or not finite-temperature corrections to the Lagrangian $ \mathcal{L} $ from Eq.~\eqref{eq:L} actually take the form $ \mathcal{L}_{\bar{\beta}} $ from Eq.~\eqref{eq:Lbeta}.
Below, we will see that our improved model avoids this instability in a more natural way that is connected to an underlying Lagrangian and works at zero temperature.

In the following, we will use the fiducial numerical parameters from Ref.~\cite{Berezhiani2018} for standard {\sc SFDM}, unless stated otherwise: $ \bar{\beta} = 2 $, $ \bar{\alpha} = 5.7 $, $ m = 1\,\rm{eV} $, and $ \Lambda = 0.05 \, \rm{meV} $.

\subsection{The {\sc MOND} limit problem}
\label{sec:prob:mondlimit}

On galactic scales, the field $ \theta $ plays a double role.
The first role is to exert an additional force on the baryons, $ \vec{a}_\theta = - \lambda \vec{\nabla} \theta $.
This follows from the phonon-baryon coupling $ - \lambda \theta \rho_b $ in $ \mathcal{L} $.
In the limit
\begin{align}
 \label{eq:MONDcondition}
 (\vec{\nabla} \theta)^2 \gg 2 m \hat{\mu} \,,
\end{align}
this acceleration has a {\sc MOND}-like form\footnote{If $ \bar{\beta} $ is not much larger than 1. If $ \bar{\beta} $ is much larger than 1, the condition becomes $ (\vec{\nabla} \theta)^2 \gg 2 m \hat{\mu} \bar{\beta} $.
For the numerical values we consider here, $ \bar{\beta} $ is of order 1.}
, because it satisfies the {\sc MOND}-like equation
\begin{align}
 \label{eq:MONDeq}
 \vec{\nabla} \left( |\vec{a}_\theta| \vec{a}_\theta \right) = \vec{\nabla} \left(a_0 \vec{a}_b\right) \,.
\end{align}
This condition is usually assumed to hold on galactic scales so that {\sc SFDM} reproduces the {\sc MOND} phenomenology of rotation curves.
However, it turns out that the condition Eq.~\eqref{eq:MONDcondition} is not always satisfied inside galaxies.
Concretely, the quantity $ \varepsilon \equiv (2 m \hat{\mu})/(\vec{\nabla} \theta)^2 $ that controls the {\sc MOND} limit can be written as
\begin{align}
 \label{eq:epsilon}
 \varepsilon \equiv \frac{2 m \hat{\mu}}{(\vec{\nabla} \theta)^2} = \frac{2 m^2}{\bar{\alpha}} \frac{10^{-6}}{a_0 M_{\rm{Pl}}} \left(\frac{a_0}{10 |\vec{a}_b|}\right) \left(10^7 \frac{\hat{\mu}}{m}\right) \approx 0.8 \cdot \left(\frac{a_0}{10 |\vec{a}_b|}\right) \left(10^7 \frac{\hat{\mu}}{m}\right) \,,
\end{align}
where we assumed a {\sc MOND}ian acceleration $ |\vec{a}_\theta| = \sqrt{a_0 |\vec{a}_b|} $ and we used the fiducial numerical parameters from Ref.~\cite{Berezhiani2018}.
We introduced a factor $ 10^7 $ in front of $ \hat{\mu}/m $ because $ 10^{-7} $ is a typical value of $ \hat{\mu}/m $ on galactic scales.
Indeed, a rough estimate is $ \hat{\mu}/m \approx |\phi_{\rm{N}}| $.
Similarly, we introduced a factor of $ 1/10 $ in front of $ a_0/|\vec{a}_b| $ because we expect the {\sc MOND} limit to be valid for $ |\vec{a}_b| \ll a_0 $.
We see that $ \varepsilon $ can easily be of order 1 on galactic scales.
When this happens, {\sc SFDM} does not have a proper {\sc MOND} limit.

It may be possible to avoid this problem by choosing different values for $ \bar{\alpha} $ and $ m $, since $ \varepsilon $ scales as $ m^2/\bar{\alpha} $, see Eq.~\eqref{eq:epsilon}.
However, a small ratio $ m^2/\bar{\alpha} $ is in tension with the other role that $ \theta $ plays on galactic scales, namely to carry the superfluid's energy density.
This role is discussed in more detail in the next section.
Here, it suffices to note that the energy density of the superfluid scales as \cite{Berezhiani2018}
\begin{align}
 \rho_{\rm{DM}} \propto \frac{m^2}{\bar{\alpha}} M_{\rm{Pl}} |\vec{a}_\theta|
\end{align}
in the {\sc MOND} limit with small $ \varepsilon $.
Thus, if we make $\varepsilon$ significantly smaller through $m^2/\bar{\alpha}$ so that $\theta$ mediates a {\sc MOND}ian force $a_{\theta} = \sqrt{a_0 a_b}$, the superfluid density also becomes significantly smaller.
But the superfluid density cannot be too small because it is needed to get strong lensing right.
This is because we cannot couple the {\sc MOND}-like phonon force to photons to satisfy the constraints from gravitational waves with an electromagnetic counterpart \cite{Boran2018, Sanders2018, Hossenfelder2019}.
Thus, one cannot fix the {\sc MOND} limit by simply choosing different numerical parameters.

\begin{figure}
 \centering
 \includegraphics[width=.49\textwidth]{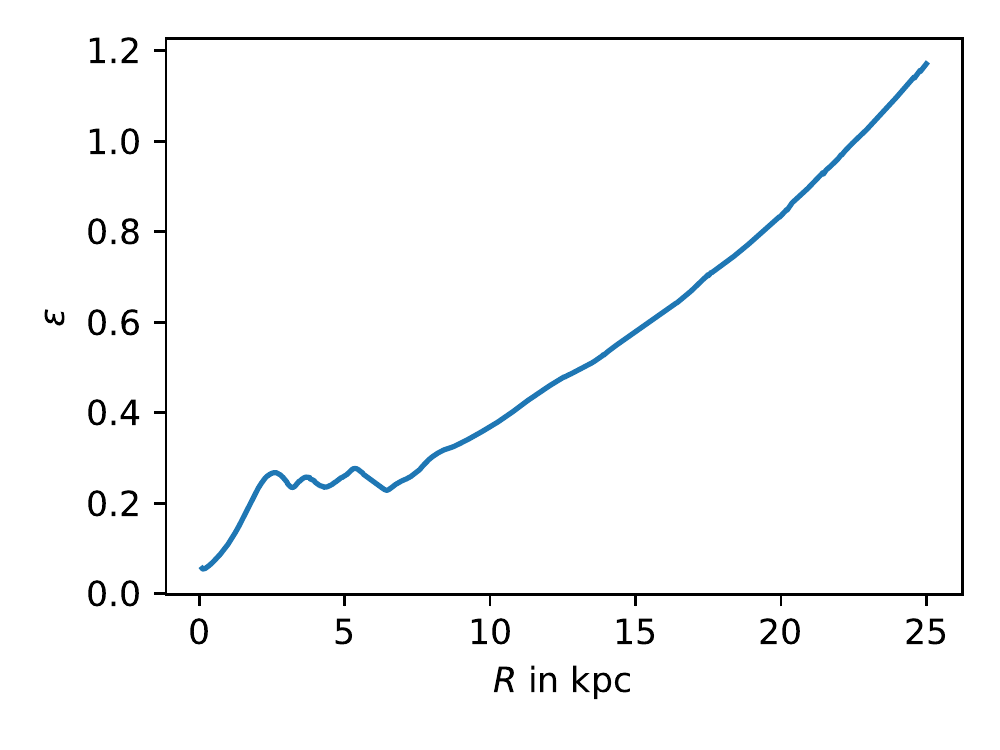}
 \includegraphics[width=.49\textwidth]{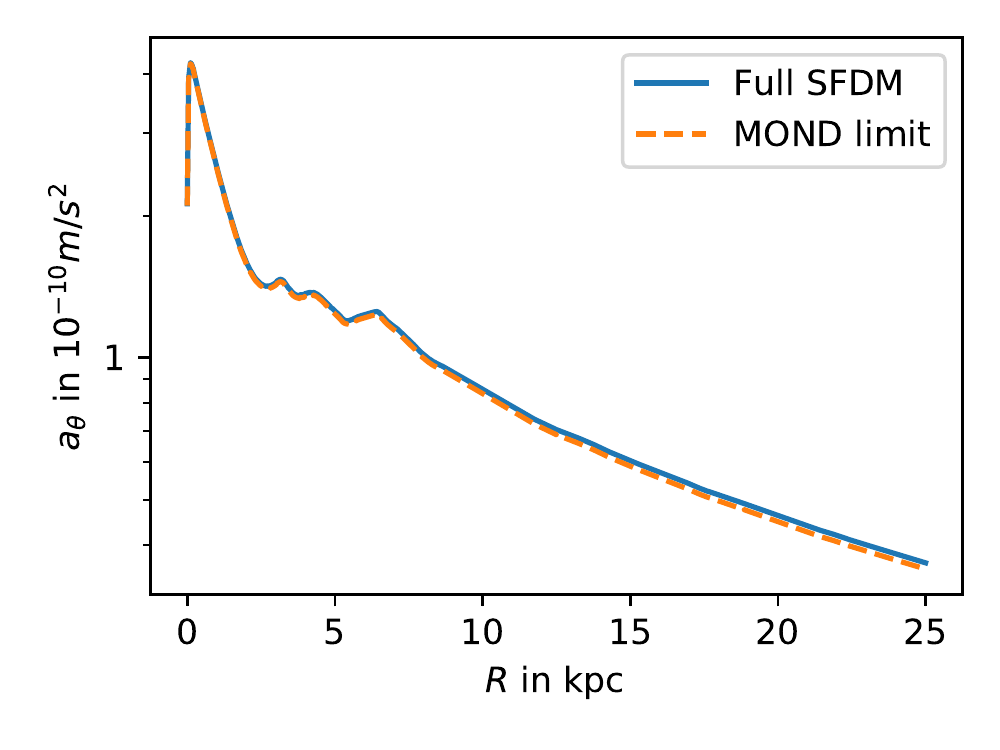}
 \caption{
     Left: The quantity $ \varepsilon = (2m\hat{\mu})/(\vec{\nabla} \theta)^2 $ that controls the idealized {\sc MOND} limit of {\sc SFDM} at $z = 0$ for the Milky Way model from Ref.~\cite{Hossenfelder2020} with $ f_b = 0.8 $.
     Right: The phonon acceleration $\vec{a}_\theta$ at $ z = 0 $ for the full SFDM model (solid blue line) and for the idealized {\sc MOND} limit from Eq.~\eqref{eq:MONDeq} (dashed orange line).
}
 \label{fig:MONDregime}
\end{figure}

The behavior of $ \varepsilon $ is illustrated in Fig.~\ref{fig:MONDregime}, left, where we plot the quantity $ \varepsilon \equiv (2 m \hat{\mu})/(\vec{\nabla} \theta)^2 $ over the cylindrical radius $ R $ for the Milky Way at $ z = 0 $.
This plot is for the {\sc SFDM} model of the Milky Way rotation curve from Ref.~\cite{Hossenfelder2020} with the parameters $ f_b = 0.8 $, $ \mu_\infty/m = 1.25 \cdot 10^{-8} $, and $r_\infty = 100\,\rm{kpc}$.
Here, $ f_b $ controls the amount of baryonic mass and $ \mu_\infty $ is a boundary condition imposed at $r_\infty$ that controls the superfluid density, see Ref.~\cite{Hossenfelder2020} for details.
These parameters give a reasonable fit of the Milky Way rotation curve.
This result is obtained by numerically solving the full {\sc SFDM} equations, it does not use the idealized {\sc MOND} limit from Eq.~\eqref{eq:MONDeq}.
Fig.~\ref{fig:MONDregime}, left, shows that $ \varepsilon $ is larger than 1 for $ \gtrsim 20\,\rm{kpc} $.
Even at $ R < 20\,\rm{kpc} $, $ \varepsilon $ is only moderately small.
This confirms our expectation from the estimate Eq.~\eqref{eq:epsilon}.

However, there is also a puzzle here.
Namely, because of the large value of $ \varepsilon $, we would expect the {\sc MOND}ian equation Eq.~\eqref{eq:MONDeq} to be a bad approximation for the phonon acceleration $ \vec{a}_\theta $.
In particular, we would expect the {\sc SFDM} rotation curve to deviate significantly from the idealized {\sc MOND} limit represented by Eq.~\eqref{eq:MONDeq}.
But if we calculate the Milky Way's rotation curve using this idealized {\sc MOND} limit, we find that the rotation curve is almost the same as for the full {\sc SFDM} equations for $ R \lesssim 25\,\rm{kpc} $.
This is illustrated in Fig.~\ref{fig:MONDregime}, right.
This is the puzzle:
How can the condition $ \varepsilon \ll 1 $ be violated but the rotation curve still be so close to the {\sc MOND}ian rotation curve?
The answer is that a numerical coincidence brings the rotation curve of the full {\sc SFDM} equations close to the rotation curve from the idealized {\sc MOND} limit.
For example, this coincidence does not work out so favorably if we choose $ \bar{\beta} = 2.5 $ instead of $ \bar{\beta} = 2 $.
This is discussed in more detail in Appendix~\ref{sec:mondregimecoincidence}.
The details of how this numerical coincidence comes about are not important for our further discussion.

Note that having a proper {\sc MOND} limit is not only about the rotation curve of an isolated galaxy.
It is also about the External Field Effect ({\sc EFE}) \cite{Milgrom1983a, Famaey2012}.
In {\sc SFDM}, it is usually assumed that there is a standard {\sc MOND}ian {\sc EFE} inside the superfluid core of a galaxy \cite{Berezhiani2018}.
However, this is true only if $\varepsilon \ll 1$, because then the $\theta$ equation of motion has the standard {\sc MOND}ian form.
If $\varepsilon$ is not small, there is no standard {\sc MOND}ian {\sc EFE} in the superfluid core -- even if the rotation curve is {\sc MOND}ian due to the numerical coincidence discussed in Appendix~\ref{sec:mondregimecoincidence}.
This is important for satellite galaxies and globular clusters of the Milky Way which are expected to be affected by this {\sc EFE} in {\sc SFDM} \cite{Berezhiani2018}.

Below, we will see that our improved model allows for a proper {\sc MOND} limit much more generally and without relying on numerical coincidences.

\subsection{The equilibrium problem}
\label{sec:prob:equilibrium}

As already mentioned in the previous section, the second role that $ \theta $ plays on galactic scales is that it carries the superfluid's energy density.
That is, it represents the superfluid in equilibrium with chemical potential $ \mu $.
This energy density $ \rho_{\rm{DM}} $ provides an additional term to the total energy density that sources the Newtonian gravitational potential $ \phi_{\rm{N}} $ \cite{Berezhiani2018},
\begin{align}
 \label{eq:rhoDMstandard}
 \rho_{\rm{DM}} = \frac{2\sqrt{2}}{3} m^{5/2} \Lambda \, \frac{3(\bar{\beta}-1) \hat{\mu} + (3-\bar{\beta}) \frac{(\vec{\nabla} \theta)^2}{2m}}{\sqrt{ (\bar{\beta}-1) \hat{\mu}  + \frac{(\vec{\nabla} \theta)^2}{2m} }} \,.
\end{align}
The additional Newtonian gravitational pull due to this energy density is negligible in the inner parts of galaxies.
There, the {\sc MOND}-like phonon force dominates.
But at large radii, the additional gravitational pull from $ \rho_{\rm{DM}} $  becomes important and is responsible for the strong lensing signal of galaxies.

That $ \theta $ plays this double role leads to a potential problem with the superfluid's equilibrium:
On the one hand, a perfect equilibrium with a chemical potential $ \mu $ requires an associated conserved quantity.
Here, this is the conserved quantity associated with a shift symmetry of $ \theta $.
On the other hand, that $ \theta $ exerts a {\sc MOND}-like force on the baryons implies that the $ \theta $ shift symmetry is broken.
Indeed, the coupling $ - \lambda \theta \rho_b $ explicitly breaks this symmetry.
This may not be a problem if $ \lambda $ is small enough so that there is at least an approximate equilibrium valid for times much longer than the timescales of interest.
Ref.~\cite{Mistele2019} estimates that such an approximate equilibrium could be valid on timescales much smaller than $ t_Q $,
\begin{align}
 \label{eq:tQ}
 t_Q^{-1} \approx \lambda \, m \frac{M_b}{M_{\rm{DM}}} \,,
\end{align}
where $ M_b $ is the total baryonic mass and $ M_{\rm{DM}} $ is the total superfluid mass.
With the fiducial parameter values from Ref.~\cite{Berezhiani2018} ($ \lambda \approx 10^{-31} $ and $ m = 1 \,\rm{eV} $), this gives $ t_Q \approx 10^8\,\mathrm{yr} \, (M_{\rm{DM}}/M_b) $.
A typical dynamical timescale of galaxies is $ 10^8\,\rm{yr} $.
Thus, the timescale $ t_Q $ is not necessarily much larger than the dynamical timescale of a galaxy so that assuming an equilibrium with chemical potential $ \mu $ may not be valid on galactic scales.
Indeed, the global estimate Eq.~\eqref{eq:tQ} concerns a galaxy as a whole and may be too optimistic.
Ref.~\cite{Mistele2019} also estimates a local timescale $ t_{\rm{loc}} $ separately for each point in space and finds that the approximate equilibrium may be valid only on timescales much shorter than $ t_{\rm{loc}}^{-1} = \lambda \, m \, (\rho_b/\rho_{\rm{DM}}) $.
In the inner parts of galaxies, $ t_{\rm{loc}} $ is typically much shorter than $ 10^8\,\rm{yr} $ since $ \rho_{\rm{DM}} \ll \rho_b $, which makes the equilibrium problem more pressing.
In principle, $ t_{\rm{loc}} $ and $ t_Q $ can be made large by making $\lambda m / \rho_{\rm{DM}} $ small.
However, in the {\sc MOND} limit, we have $\lambda m / \rho_{\rm{DM}} \propto M_{\rm{Pl}}^{-1} (a_0/a_b)^{1/4} \rho_{\rm{DM}}^{-1/2} $.
Thus, at fixed $ a_0 $ (to keep the phonon force fixed) and at fixed $ \rho_{\rm{DM}} $ (to keep the energy density fixed), there is no freedom to adjust $\lambda m /\rho_{\rm{DM}}$.
Below, we will see that our improved model has a perfect equilibrium that is valid on timescales much larger than $ t_Q $ or $ t_{\rm{loc}} $.

\section{An improved {\sc SFDM} model}
\label{sec:improved}

Above, we have seen that the equilibrium problem and the {\sc MOND} limit problem are consequences of the double role that the phonon field $ \theta $ plays in standard {\sc SFDM}:
$ \theta $ carries both the additional {\sc MOND}-like force and the superfluid's energy density.
This leads to the main idea behind our improved model, which is to disentangle the two roles that $ \theta $ plays in standard {\sc SFDM} by assigning each role to a separate field.
For brevity, we will refer to this model as two-field {\sc SFDM}.

\subsection{The model}
\label{sec:model}

We start with the {\sc EFT} from Eq.~\eqref{eq:L},
\begin{align}
 \mathcal{L} = f(K_\theta - m^2) - \lambda \, \theta \, \rho_b \,.
\end{align}
We will modify this Lagrangian to avoid the problems discussed in the previous section.
First, we introduce two fields $ \theta_+ $ and $ \theta_- $.
The role of $ \theta_- $ will be to carry the superfluid's equilibrium chemical potential and the role of $ \theta_+ $ will be to carry the additional {\sc MOND}ian force.
Then, we postulate the Lagrangian
\begin{align}
 \label{eq:Limpstart}
 \mathcal{L}_{\rm{imp}} = f(K_+ + K_- - m^2) - \lambda \, \theta_+ \, \rho_b \,,
\end{align}
where $K_\pm$ are defined analogously to $K_\theta$,
\begin{subequations}
\begin{align}
 K_+ &= \nabla^\alpha \theta_+ \nabla_\alpha \theta_+ \,, \\
 K_- &= \nabla^\alpha \theta_- \nabla_\alpha \theta_- \,.
\end{align}
\end{subequations}
Here, we couple only $ \theta_+ $ to the baryon density, not $ \theta_- $.
Thus, this Lagrangian is shift-symmetric in $ \theta_- $ and there is an associated conserved quantity $Q$.
In equilibrium, we can then introduce a chemical potential $\mu$ by replacing the Hamiltonian $H$ with $H - \mu \, Q$.
On the level of the Lagrangian, this corresponds to a shift of $\dot{\theta}_-$ \cite{Mistele2019, Kapusta1981, Haber1982, Bilic2008},
\begin{align}
 \label{eq:mu-}
 \dot{\theta}_- \to \dot{\theta}_-  + \mu \,.
\end{align}
Equivalently, we could consider solutions $\theta_- = \mu \, t$ with otherwise time-independent fields to find the same physical equilibrium.
However, introducing $\mu$ as in Eq.~\eqref{eq:mu-} makes it clear that $\mu$ is a chemical potential in the statistical physics sense.
Since the $\theta_-$ shift symmetry is exact, $Q$ is conserved exactly.
In contrast, the analogous quantity in standard {\sc SFDM} is conserved only approximately on timescales much shorter than $t_Q$.
Thus, the equilibrium problem from Sec.~\ref{sec:prob:equilibrium} is absent in two-field {\sc SFDM}:
The equilibrium with chemical potential $\mu$ is not limited to timescales shorter than $t_Q$.

Consider now a non-relativistic equilibrium with chemical potential $ \mu = m + \mu_{\rm{nr}} $ with $ |\mu_{\rm{nr}}| \ll m $.
Then, the $\theta_-$ equation of motion is solved by setting $\theta_-=0$ (that is, $\theta_-=0$ after the shift $ \dot{\theta}_- \to \dot{\theta}_- + \mu $).
In this case, the equation of motion for a time-independent $ \theta_+ $ is exactly the same as the equation of motion for a time-independent $ \theta $ from standard {\sc SFDM} with the same chemical potential,
\begin{align}
 \label{eq:eomth+}
 \vec{\nabla} \left[2 f'\left((\vec{\nabla} \theta_+)^2 - 2m \hat{\mu}\right) \vec{\nabla} \theta_+ \right] = \lambda \, \rho_b \,,
\end{align}
where, as in standard {\sc SFDM}, $ \hat{\mu} = \mu_{\rm{nr}} - m \phi_{\rm{N}} $.
The energy density derived from $ \mathcal{L}_{\rm{imp}} $ in this limit also matches that derived from standard {\sc SFDM}.
Thus, the Lagrangian $ \mathcal{L}_{\rm{imp}} $ reproduces the basic phenomenology of standard {\sc SFDM} on galactic scales but avoids the equilibrium problem.

However, so far $\mathcal{L}_{\rm{imp}}$ does not address the stability problem (see Sec.~\ref{sec:prob:stability}) and the {\sc MOND} limit problem (see Sec.~\ref{sec:prob:mondlimit}).
We will now see that both problems can be solved by one more modification of $ \mathcal{L}_{\rm{imp}} $.
Consider first the stability problem.
The second-order Lagrangian for perturbations $ \delta_\pm $ of $ \theta_\pm$ around background fields $\theta^0_\pm$ reads
\begin{align}
\label{eq:Limppert}
\begin{split}
 \mathcal{L}_{\rm{imp},\rm{pert}}
   = &\left[f_0' \, g^{\alpha \beta} + 2 f_0'' \, \nabla^\alpha \theta^0_+ \nabla^\beta \theta^0_+ \right] \nabla_\alpha \delta_+ \nabla_\beta \delta_+  + \\
    &\left[f_0' \, g^{\alpha \beta} + 2 f_0'' \, \nabla^\alpha \theta^0_- \nabla^\beta \theta^0_- \right] \nabla_\alpha \delta_- \nabla_\beta \delta_- + \\
    &\left[4 f_0'' \, \nabla^\alpha \theta^0_+ \nabla^\beta \theta^0_- \right] \nabla_\alpha \delta_+ \nabla_\beta \delta_- \,.
\end{split}
\end{align}
For equilibrium solutions on galactic scales with $ K^0_+ + K^0_- - m^2 < 0 $,
\begin{subequations}
 \label{eq:f0imp1}
\begin{align}
  f_0' &= \Lambda \sqrt{|K^0_+ + K^0_- - m^2|} > 0 \,, \\
 f_0'' &= - \frac{\Lambda}{2} \frac{1}{\sqrt{|K^0_+ + K^0_- - m^2|}} < 0 \,.
\end{align}
\end{subequations}
Here, $ K^0_\pm $ means $ K_\pm $ evaluated at the background solution $ \theta_\pm^0 $.
The first two lines in Eq.~\eqref{eq:Limppert} are two copies of the perturbed Lagrangian $\mathcal{L}_{\rm{pert}}$ from standard {\sc SFDM} from Eq.~\eqref{eq:Lpert}.
The third line comes from the mixing of $ \theta_+ $ and $ \theta_- $ in $ f(K_+ + K_- - m^2) $.

Consider now non-relativistic equilibrium solutions with chemical potential $\mu_0$, i.e. consider $\dot{\theta}^0_- = \mu_0$ with $ |g^{00}\mu_0^2 - m^2| \ll m^2$ and time-independent $ \theta^0_+ $.
In this limit, $ f_0' $ and $ f_0'' $ in Eq.~\eqref{eq:f0imp1} are the same as $ f_0' $ and $ f_0'' $ in Eq.~\eqref{eq:f0} in the analogous limit.
This is why we reuse the same symbols here.
Specifically,
\begin{subequations}
\label{eq:f0imp}
\begin{align}
  f_0' &= \Lambda \sqrt{|(\vec{\nabla} \theta^0_+)^2 - 2 m \hat{\mu}_0|} \,, \\
 f_0'' &= - \frac{\Lambda}{2} \frac{1}{\sqrt{|(\vec{\nabla} \theta^0_+)^2 - 2 m \hat{\mu}_0|}} \, .
\end{align}
\end{subequations}
Such solutions are stable if the Hamiltonian $ \mathcal{H}_{\rm{imp,pert}} $ associated with $ \mathcal{L}_{\rm{imp,pert}} $ is bounded from below for such background solutions.
We now show that this is not the case, since $\mathcal{L}_{\rm{imp}}$ inherits the instability problem of standard {\sc SFDM}.
Then, we investigate how to fix this problem.

In the limit under consideration, the third line in $ \mathcal{L}_{\rm{imp,pert}} $ reads:
\begin{align}
 - 4 f_0'' \vec{\nabla} \theta^0_+ \, \mu_0 \, \vec{\nabla} \delta_+ \, \dot{\delta}_- \,.
\end{align}
This is first order in time derivatives of the perturbations and does therefore not contribute to the Hamiltonian.
The first two lines then give the Hamiltonian
\begin{align}
 \label{eq:Himppert}
 \begin{split}
 \mathcal{H}_{\rm{imp,pert}}
 = & \left(f'_0 g^{00} + 2 (g^{00})^2 f_0'' (\dot{\theta}^0_+)^2 \right) \dot{\delta}_+^2      + \left(f'_0 (\vec{\nabla} \delta_+)^2 - 2 f''_0 (\vec{\nabla} \theta^0_+ \vec{\nabla} \delta_+)^2  \right) + \\
   & \left(f'_0 g^{00} + 2 (g^{00})^2 f_0'' (\dot{\theta}^0_-)^2 \right) \dot{\delta}_-^2 + \left(f'_0 (\vec{\nabla} \delta_-)^2 - 2 f''_0 (\vec{\nabla} \theta^0_- \vec{\nabla} \delta_-)^2  \right) \,.
\end{split}
\end{align}
Since $ f_0' > 0 $ and $ f_0'' < 0 $, the only terms that can be negative in this Hamiltonian are the prefactors of $ \dot{\delta}_-^2 $ and $ \dot{\delta}_+^2 $.
Further, we have $ \dot{\theta}^0_+ = 0 $ and $ \dot{\theta}^0_- = \mu_0 $ for the equilibrium solutions under consideration.
This leaves only the prefactor of $ \dot{\delta}_-^2 $.
This prefactor is $ g^{00} f_0' + 2 (g^{00})^2 f_0''\mu_0^2 $.
This is exactly the same as the prefactor of $ \dot{\delta}^2 $ that is responsible for the instability in standard {\sc SFDM} discussed in Sec.~\ref{sec:prob:stability}.
Thus, our improved Lagrangian suffers from a very similar instability.
However, it is now much easier to fix this problem.
Indeed, consider an additional term proportional to $ K_- $ in the Lagrangian:
\begin{align}
 \Delta \mathcal{L}_{\rm{imp}} = A K_- \,,
\end{align}
where $ A > 0 $ does not depend on $ \theta_+ $ and derivatives, but is otherwise allowed to depend on other fields.
This term does not affect the background solutions for $ \theta_- $ and $ \theta_+ $ (for a given $\phi_{\rm{N}}$, see below).
The reason is that it does not affect the $\theta_+$ equation of motion by construction and it adds a term $ \nabla_\alpha (A \nabla^\alpha \theta_-) $ to the $ \theta_- $ equation of motion, which vanishes for $\dot{\theta}_- = \mu$ and time-independent $ A $, as we assume in equilibrium.
However, it does affect the stability of perturbations.
Namely, $ \mathcal{H}_{\rm{imp,pert}} $ gains two additional terms,
\begin{align}
\Delta \mathcal{H}_{\rm{imp,pert}} = A \dot{\delta}_-^2 + A (\vec{\nabla} \delta_-)^2\,.
\end{align}
This makes the solutions under consideration stable if
\begin{align}
 \label{eq:stability}
 A > -\left(f_0' g^{00} + 2 (g^{00})^2 f_0'' \mu_0^2\right) \,.
\end{align}
Thus, such a term can fix the stability problem while keeping the background solutions for $ \theta_+ $ and $ \theta_- $ the same.
Note that a term $ A K_\theta $ in standard {\sc SFDM} could similarly fix the stability problem there.
However, this would significantly change the {\sc MOND}ian force in standard {\sc SFDM}.
The reason is that $A K_\theta$ contains spatial derivatives of the field $\theta$ which carries this force.
In contrast, the $ A K_- $ term in our two-field model does not affect the force carried by $ \theta_+ $.

This leaves us with the {\sc MOND} limit problem.
That is, in standard {\sc SFDM}, the condition $ (\vec{\nabla} \theta)^2 \gg 2 m \hat{\mu} $ is not always fulfilled in galaxies so that the force carried by $\theta$ is not always {\sc MOND}ian.
So far, the analogous condition $ (\vec{\nabla} \theta_+)^2 \gg 2 m \hat{\mu} $ in our two-field model has the exact same problem.
As discussed in Sec.~\ref{sec:prob:mondlimit}, this could be fixed by making the ratio $ m^2/\bar{\alpha} $ smaller.
But then the superfluid's energy density in galaxies becomes smaller by roughly the same factor.
This is problematic since the superfluid's energy density is important to get enough strong lensing.
However, in our two-field model, another way to fix this problem is through the $A K_-$ term.
Namely, this term adds an additional term of order $ A m^2 $ to the superfluid's energy density.
Then, we may hope to fix the {\sc MOND} limit problem, if we make the ratio $m^2/\bar{\alpha}$ smaller to satisfy $ (\vec{\nabla} \theta_+)^2 \gg 2 m \hat{\mu} $, but keep the energy density roughly the same through the $A m^2$ term.

To make this concrete, we consider one possible origin of the $ A K_- $ term.
We take $ \theta_- $ to be the phase of a complex scalar field $ \phi_- = \rho_- e^{-i \theta_-} / \sqrt{2} $ and
add to $ \mathcal{L}_{\rm{imp}} $ the standard Lagrangian for a superfluid with a quartic interaction \cite{Schmitt2015, Sharma2019}:
\begin{align}
\begin{split}
 \mathcal{L}_- &= (\nabla_\alpha \phi_-)^* (\nabla^\alpha \phi_-) - m^2 |\phi_-|^2 - \lambda_4 |\phi_-|^4 \\
               &= \frac12 \left( K_{\rho_-} + \rho_-^2 (K_- - m^2) \right) - \frac{\lambda_4}{4} \rho_-^4 \,.
\end{split}
\end{align}
Here, $ m $ is the mass, $ \lambda_4 $ is the quartic coupling constant, and
\begin{subequations}
\begin{align}
 K_{\rho_-} &= \nabla^\alpha \rho_- \nabla_\alpha \rho_- \,, \\
        K_- &= \nabla^\alpha \theta_- \nabla_\alpha \theta_- \,.
\end{align}
\end{subequations}
If we introduce a chemical potential $ \mu $ by shifting $ \dot{\theta}_- \to \dot{\theta}_- + \mu $ and neglect derivatives of $ \rho_- $, the equation of motion for $ \rho_- $ has the solution
\begin{align}
 \rho_-^2 = \frac{g^{00} \mu^2 - m^2}{\lambda_4} \,,
\end{align}
if $ g^{00} \mu^2 > m^2 $.
This is the usual superfluid solution.
In the non-relativistic limit with $ \mu = m + \mu_{\rm{nr}} $ with $ |\mu_{\rm{nr}}| \ll m $ and $ g^{00} = 1 - 2 \phi_{\rm{N}} $, this is
\begin{align}
 \rho_-^2 = \frac{2 m (\mu_{\rm{nr}} - m \phi_{\rm{N}})}{\lambda_4} = \frac{2 m \hat{\mu}}{\lambda_4} \,,
\end{align}
and the dominant contribution to the energy density is
\begin{align}
 \rho_{\rm{DM}-} = m^2 \rho_-^2 = \frac{2m^3 \hat{\mu}}{\lambda_4} \,.
\end{align}
In this model, we have
\begin{align}
 A = \frac{\rho_-^2}{2} \,.
\end{align}
For the standard equilibrium superfluid, this is $ A = m \hat{\mu} / \lambda_4 $.
The total dark matter density then has two terms.
One term from the previous $ \mathcal{L}_{\rm{imp}} $ and the new term of order $ A m^2 $,
\begin{subequations}
\label{eq:rhoDM}
\begin{align}
   \rho_{\rm{DM}} &= \rho_{\rm{DM}-} + \rho_{\rm{DM}+} \,, \\
 \rho_{\rm{DM}-} &= \frac{2 m^3 \hat{\mu}}{\lambda_4} \,, \\
 \rho_{\rm{DM}+} &= 2 m^2 \Lambda \, \sqrt{(\vec{\nabla} \theta_+)^2 - 2 m \hat{\mu}} \,.
\end{align}
\end{subequations}
If we make $ m^2/\bar{\alpha} $ small to fix the {\sc MOND} limit problem, $ \rho_{\rm{DM}+} $ becomes small.
However, $ \rho_{\rm{DM}-} $ can take over the role of $ \rho_{\rm{DM}+} $, since it scales differently.
Concretely, we can estimate $ \hat{\mu} \approx m |\phi_{\rm{N}}| $.
Then, we see that we need to keep $ m^4/\lambda_4 $ large enough to produce enough strong lensing while making $ m^2/\bar{\alpha} $ small enough to fix the {\sc MOND} limit problem.
The stability criterion Eq.~\eqref{eq:stability} now reads:
\begin{align}
 \frac{m \hat{\mu}}{\lambda_4} > \frac{\Lambda m^2}{\sqrt{|K_+ + K_- - m^2|}} \left(1 - \frac{|K_+ + K_- - m^2|}{m^2}\right) \approx \frac{\Lambda m^2}{\sqrt{|(\vec{\nabla} \theta_+)^2 - 2 m \hat{\mu}|}} \,.
\end{align}
If we make $ m^2/\bar{\alpha} $ small enough such that $ (\vec{\nabla} \theta_+)^2 \gg 2 m \hat{\mu} $, this further simplifies to
\begin{align}
 \label{eq:stabilitya}
 |\vec{a}_+| > a_0\,  \frac{\lambda_4}{\bar{\alpha}^2} \, \frac{m}{\hat{\mu}} \equiv a_{+\rm{min}} \,,
\end{align}
where $ \vec{a}_+ = - \lambda \vec{\nabla} \theta_+ $ is the acceleration of the baryons due to $\theta_+$.
This implies a minimum acceleration $ a_{+\rm{min}} $ due to $ \theta_+ $ below which the equilibrium solutions under consideration become unstable.
In the {\sc MOND} limit with $ |\vec{a}_+| = \sqrt{a_0 |\vec{a}_b|} $, this is
\begin{align}
 \label{eq:stabilitya2}
 a_{\rm{b}} \left(10^7 \frac{\hat{\mu}}{m}\right)^2 > a_0 \left(10^7 \frac{\lambda_4}{\bar{\alpha}^2}\right)^2 \equiv \bar{a} \,.
\end{align}
Here, we introduced a factor of $ 10^7 $ because $ 10^{-7} $ is a typical value of $ \hat{\mu}/m $ on galactic scales.
Thus, stability typically requires $ a_b > \bar{a} $, but the precise condition depends on $\hat{\mu}/m$ and is therefore different in different galaxies and at different radii.
For a {\sc MOND}ian phenomenology, we want to keep the equilibrium solutions valid for baryonic accelerations well below $ a_0 $.
Thus, we need a very small ratio $ \lambda_4/\bar{\alpha}^2 $.\footnote{
    To stay away from $ K_+ + K_- - m^2 = 0 $, where the {\sc SFDM} {\sc EFT} is ill-defined, we need $ (\vec{\nabla} \theta_+)^2 > 2 m \hat{\mu} $.
    This gives another minimum acceleration.
    However, in the {\sc MOND} limit, we have $ (\vec{\nabla} \theta_+)^2 \gg 2 m \hat{\mu} $.
    Therefore, we will assume that the minimum acceleration from stability concerns discussed above is more important on galactic scales.
    A third type of minimum acceleration is discussed in Appendix~\ref{sec:origin}.
}

To summarize, our two-field {\sc SFDM} model is
\begin{align}
\label{eq:Limp}
\begin{split}
 \mathcal{L}_{\rm{imp}} &= \mathcal{L}_- + f(K_+ + K_- - m^2) - \lambda \, \theta_+\, \rho_b \\
                        &= \frac12 \left( K_{\rho_-} + \rho_-^2 (K_- - m^2) \right) - \frac{\lambda_4}{4} \rho_-^4 + f(K_+ + K_- - m^2) - \lambda \, \theta_+\, \rho_b \,.
\end{split}
\end{align}
where $ \theta_+ $ carries the additional {\sc MOND}-like force and $ \phi_- = e^{-i\theta_-} \rho_- / \sqrt{2} $ carries the superfluid's energy density.
It has four free parameters $ m $, $ \bar{\alpha} $, $ \Lambda $, and $ \lambda_4 $.
This is the same number of free parameters as in standard {\sc SFDM}, where the free parameters are $ m $, $ \bar{\alpha} $, $ \Lambda $, and $ \bar{\beta} $.\footnote{In principle, $ \bar{\beta} $ is not a free parameter since the finite-temperature corrections it parametrizes should follow from the zero-temperature Lagrangian from Eq.~\ref{eq:L}. However, for practical purposes, it is a free parameter, as discussed in Sec.~\ref{sec:prob:stability}.}
We have shown how our model can avoid the instability problem (see Sec.~\ref{sec:prob:stability}), the {\sc MOND} limit problem (see Sec.~\ref{sec:prob:mondlimit}), and the equilibrium problem (see Sec.~\ref{sec:prob:equilibrium}).
Numerically, we need $\lambda_4/\bar{\alpha}^2 \ll 10^{-7}$ to avoid the stability problem.
To avoid the {\sc MOND} limit problem, we need $m^2/\bar{\alpha}$ much smaller than in standard {\sc SFDM} and $m^4/\lambda_4$ large enough to give enough lensing.
In Secs.~\ref{sec:params} and \ref{sec:pheno}, we will choose concrete numerical values and demonstrate some galactic phenomenology of the model.
We will further discuss the qualitative interpretation of our model in Sec.~\ref{sec:discussion}.

As in the standard {\sc SFDM} {\sc EFT}, we have so far taken the form of the function $ f $ as given.
Ref.~\cite{Berezhiani2015} discusses a possible origin of $ f $ from an underlying Lagrangian that is valid also at $ K - m^2 = 0 $, where the {\sc EFT} is ill-defined.
In Appendix~\ref{sec:origin}, we show that it is straightforward to extend this mechanism to two-field {\sc SFDM}.

\subsection{Sound speed}
\label{sec:soundspeed}

Low-energy perturbations on top of non-relativistic equilibrium superfluids usually have a single low-energy mode with sound speed $ c_s^2 = \mathcal{O}(\hat{\mu}/m) \ll 1 $.
In our model, there are two low-energy modes.
Roughly, these correspond to the two fields $ \theta_+ $ and $ \theta_- $.
If we neglect the mixing between $ \theta_+ $ and $ \theta_- $ from the $ f(K_+ + K_- - m^2) $ term in our Lagrangian, we find that a mix of $\theta_-$ and $\rho_-$ gives a gapless mode with the usual phonon dispersion relation of a superfluid with quartic interactions, i.e. a sound speed
\begin{align}
 \label{eq:cs-}
 c_{s-}^2 = \frac{\hat{\mu}}{m} \,.
\end{align}
In contrast, the $ \theta_+ $ perturbations are superluminal,
\begin{align}
 \label{eq:cs+}
 c_{s+}^2 = 1 + \gamma^2 \,,
\end{align}
where $ \gamma $ is the cosine of the angle between the wave vector of the perturbation and the background $ \vec{\nabla} \theta_+ $.
This is the usual result for {\sc RAQUAL}-type Lagrangians in the deep-{\sc MOND} limit \cite{Bekenstein1984, Bruneton2007b}.
In Appendix~\ref{sec:soundspeed:calculation}, we show that corrections to Eqs.~\eqref{eq:cs-} and \eqref{eq:cs+} from mixing of $ \theta_+ $ and $ \theta_- $ are typically small on galactic scales.

This superluminality does not necessarily imply problems with causality, as discussed in Refs.~\cite{Bruneton2007, Bruneton2007b, Babichev2008}.\footnote{
    It could, however, indicate the lack of a conventional Wilsonian UV completion \cite{Adams2006, Babichev2018}.
    In this case, a non-Wilsonian UV completion like classicalization may be possible \cite{Dvali2011, Dvali2012, Vikman2013, Addazi2019}.
}
Still, we can make $c_{s+}$ subluminal by replacing $f(K_+ + K_- -m^2)$ in our Lagrangian with
\begin{align}
 f\left(K_+ + K_- -m^2 + C (\nabla_\alpha \theta_+ \nabla^\alpha \theta_-)^2\right) \,,
\end{align}
with a suitable constant $C$.
This change does not affect the equilibrium equations of motion since all additional terms are proportional to $\nabla_\alpha \theta_+ \nabla^\alpha \theta_-$ which vanishes in equilibrium.
However, perturbations around equilibrium are affected.
If we choose $Cm^2 = \mathcal{O}(1)$ with $Cm^2 \geq 1$, it is straightforward to show that $c_{s+}$ becomes subluminal, $ c_{s+}^2 = (1 + \gamma^2)/(1+C \mu^2)$.
Other effects on our stability analysis and sound speed calculation are negligible.
Below, we leave out such a term for simplicity.

Since $ \theta_+ $ couples directly to the baryonic density $ \rho_b $ and since $ \theta_- $ mixes with $ \theta_+ $, both modes are relevant to Cherenkov radiation and binary pulsar constraints.
However, the standard {\sc SFDM} picture from galactic scales is not directly applicable in these cases.
The reason is that these constraints come from environments with accelerations much larger than $ a_0 $, where standard {\sc SFDM} phenomenology anyway cannot be valid.
Indeed, naively extending the galaxy-scale phenomenology of {\sc SFDM} to the solar system gives additional accelerations $ \vec{a}_+ $ that are much too large \cite{Berezhiani2015}.
One way to satisfy the solar system constraints could be to introduce higher-derivative terms like in Ref.~\cite{Babichev2011}.
Then, one could test Cherenkov radiation and binary pulsar constraints for the specific resulting model.
More generally, it makes sense to consider these constraints only after we understand the high-acceleration regime of {\sc SFDM}.
Studying this is left for future work.

\subsection{Halo profile in the absence of baryons}
\label{sec:halo}

In our model, Poisson's equation for the Newtonian gravitational potential $ \phi_{\rm{N}} $ on galactic scales is
\begin{align}
 \Delta \phi_{\rm{N}} = 4 \pi G \left(\rho_b + \rho_{\rm{DM}-}\right) \,,
\end{align}
where we have assumed that $ \rho_{\rm{DM}-} \gg \rho_{\rm{DM}+} $ as discussed in Sec.~\ref{sec:model}.
This can be rewritten in terms of $ \hat{\mu} = \mu_{\rm{nr}} - m \phi_{\rm{N}} $,
\begin{align}
 \Delta \left(-\frac{\hat{\mu}}{m}\right) = 4 \pi G \rho_b + \frac{m^4}{\lambda_4 M_{\rm{Pl}}^2} \left(\frac{\hat{\mu}}{m}\right) \,,
\end{align}
where we used the expression for $ \rho_{\rm{DM}-} $ from Eq.~\eqref{eq:rhoDM} and $ 8 \pi G = M_{\rm{Pl}}^{-2} $.
We now define
\begin{align}
 r_0^2 =  \lambda_4 \, \frac{M_{\rm{Pl}}^2}{m^4} \,.
\end{align}
Then, outside the baryonic energy density, the Poisson equation has solutions
\begin{align}
 \label{eq:halonobaryons}
 \hat{\mu}/m \propto \cos(r/r_0 + \delta_0) /r \,,
\end{align}
with constant $ \delta_0 $.
Thus, the dark matter profile away from the baryons is $ \rho_{\rm{DM}-} \propto \hat{\mu} \propto \cos(r/r_0 + \delta_0)/r $ with a characteristic length scale $ r_0 $.
This specific halo profile is a consequence of our choice of a $ |\phi_-|^4 $ interaction term which leads to $ \rho_{\rm{DM}-} \propto \hat{\mu}/m $.
In contrast, a $ |\phi_-|^6 $ interaction leads to $ \rho_{\rm{DM}-} \propto \sqrt{\hat{\mu}/m} $ which is closer to the energy density in standard {\sc SFDM} \cite{Berezhiani2015}.
The energy densities can be of the same order of magnitude for both types of interaction, but the shapes are different\footnote{
    For a given solution $ \rho_{\rm{DM}}(r) $, another solution is $ \xi \rho_{\rm{DM}}(r) $ for a $ |\phi_-|^4 $ interaction and without baryons.
    But for $ |\phi_-|^6 $ it is $ \xi \rho_{\rm{DM}}(r/\sqrt{\xi}) $.
    That is, multiplying the central density by $ \xi $ also multiplies the density in the $ |\phi_-|^4  $ case, but multiplies and stretches it in the $ |\phi_-|^6 $ case.
}, see Sec.~\ref{sec:pheno} and Ref.~\cite{Sharma2019} for explicit examples.
We chose the more standard quartic interaction for simplicity, but other choices are certainly possible.

Eq.~\eqref{eq:halonobaryons} holds in a zero-temperature equilibrium.
Ref.~\cite{Sharma2019} shows that assuming zero temperature is a good approximation at small and intermediate radii, but not at larger radii where the zero-temperature $\hat{\mu}/m$ reaches zero.
Thus, for a realistic galaxy in equilibrium, Eq.~\eqref{eq:halonobaryons} will be modified by the presence of baryons at small radii and by finite-temperature effects at large radii.
For the numerical parameter values we consider below (see Sec.~\ref{sec:params}), these finite-temperature effects will not be important on galactic scales.\footnote{
    Finite-temperature effects become important once $n(r) \equiv \rho_{\rm{DM}-}(r)/m$ falls below the critical value $n_c \propto m^3 (M_{\rm{DM}}/r_0)^{3/2}$, where $M_{\rm{DM}}$ is the superfluid core's mass \cite{Sharma2019}.
    In Sec.~\ref{sec:params}, we choose parameters such that $\rho_{\rm{DM}-}$ is similar as in Ref.~\cite{Sharma2019}, while $m$ is much smaller.
    Thus, $n(r)$ is much larger and $n_c$ is much smaller in our case so that finite-temperature effects are small on galactic scales.
}
Therefore, we include the effects of baryons, but no finite-temperature effects in our explicit calculations below.

\section{Choosing parameters}
\label{sec:params}

As mentioned in Sec.~\ref{sec:model}, our model has four free parameters: $ \bar{\alpha} $, $ \Lambda $, $ m $, and $ \lambda_4 $.
We will now express these in terms of four other parameters that are more directly related to phenomenology.

Our first parameter is the {\sc MOND}ian acceleration scale $ a_0 $,
\begin{align}
 a_0 = \frac{\bar{\alpha}^3 \Lambda^2}{M_{\rm{Pl}}} \,,
\end{align}
that has the same form as in standard {\sc SFDM} (see Sec.~\ref{sec:sfdm}).
Our second parameter is the characteristic length scale $ r_0 = \sqrt{\lambda_4} \, M_{\rm{Pl}}/m^2 $ in the superfluid energy density on galactic scales (see Sec.~\ref{sec:halo}).
Our third parameter is the minimum acceleration scale $ \bar{a} = a_0 (10^7 \lambda_4/\bar{\alpha}^2)^2 $ (see Sec.~\ref{sec:model}).

As our fourth parameter, we choose the non-relativistic $ 2 \to 2 $ self-interaction cross-section of $\phi_- $ particles due to the $ \lambda_4 |\phi_-|^4 $ interaction \cite{Peskin1995, Pitaevskii2003}:
\begin{align}
 \label{eq:sigmam}
 \frac{\sigma}{m} = \frac{4\pi}{2!} \frac{1}{m} \frac{(4 \lambda_4)^2}{64 \pi^2 (2m)^2} = \frac{1}{8 \pi} \frac{\lambda_4^2}{m^3} \,.
\end{align}
In Ref.~\cite{Berezhiani2015}, the self-interaction cross-section is used to estimate the size of the superfluid cores of galaxies (see also Sec.~\ref{sec:RTRNFW} below).
Its value was chosen such that constraints from cluster mergers are satisfied \cite{Berezhiani2015, Berezhiani2018}.
Below, we choose $ \sigma/m $ from Eq.~\eqref{eq:sigmam} to have the same value as in Ref.~\cite{Berezhiani2018}.
However, one should be careful with this procedure.
One reason is that, as discussed in Ref.~\cite{Berezhiani2015}, standard cluster merger constraints may need to be carefully revisited for superfluids.
Another reason is that we calculated $ \sigma/m $ only from the $ \lambda_4 |\phi_-|^4 $ term in our Lagrangian $ \mathcal{L}_{\rm{imp}} $.
Using this cross-section to satisfy constraints on the total self-interaction rate implicitly assumes that terms from the $ f(K_+ + K_- - m^2) $ part of the Lagrangian are negligible.
Estimating the effects of this term is not straightforward due to its non-standard form.
Indeed, in Ref.~\cite{Berezhiani2015}, the value of the self-interaction cross-section is simply assumed and not calculated from an underlying Lagrangian.
Investigating this in more detail is left for future work.

To sum up, we will specify our model through the {\sc MOND} acceleration scale $ a_0 $, the dark matter density characteristic length scale $ r_0 $, the minimum acceleration $ \bar{a} $, and the self-interaction cross-section $ \sigma/m $.
The parameters $ \bar{\alpha} $, $ \Lambda $, $ m $, and $ \lambda_4 $ can be expressed through these quantities:
\begin{equation}
 \begin{alignedat}{2}
 \bar{\alpha} &= 10^3 \, c_{\bar{\alpha}} \, \left(\frac{a_0}{\bar{a}}\right)^{1/4} \left(\frac{M_{\rm{Pl}}^{3} (\sigma/m)^{2}}{r_0^{3}}\right)^{1/5} \,,   \quad & m &= c_m \left(\frac{M_{\rm{Pl}}^{4} (\sigma/m)}{r_0^{4}}\right)^{1/5} \,, \\
\Lambda &= 10^{-5}  c_\Lambda \left(a_0 \bar{a}^3\right)^{1/8} \left(\frac{r_0^{9} }{M_{\rm{Pl}}^{4} (\sigma/m)^{6}}\right)^{1/10} \,, \quad &  \lambda_4 &= c_{\lambda_4} \left(\frac{M_{\rm{Pl}}^{3} (\sigma/m)^{2}}{r_0^{3}}\right)^{2/5} \,,
 \end{alignedat}
\end{equation}
where
\begin{equation}
 \begin{alignedat}{2}
  c_{\bar{\alpha}} &= 2 \cdot 2^{7/10} \sqrt{5} \, \pi^{2/5} \approx 11.48 \,, \quad &  c_m &= 2^{3/5} \, \pi^{1/5} \approx 1.91 \,, \\
 c_\Lambda &= \left(4 \cdot 2^{1/20} \cdot 5^{1/4} \, \pi^{3/5}\right)^{-1} \approx 0.08 \,,   \quad & c_{\lambda_4} &= 4 \cdot 2^{2/5} \pi^{4/5} \approx 13.19 \,.
 \end{alignedat}
\end{equation}

In the following, we use numerical values that give a phenomenology on galactic scales that is close to that of standard {\sc SFDM} with the fiducial parameters from Ref.~\cite{Berezhiani2018}.
This is useful for illustrating similarities and differences between standard {\sc SFDM} and two-field {\sc SFDM}, but different choices with different phenomenology are certainly possible.
Concretely, we choose:
\begin{equation}
 \begin{alignedat}{2}
       a_0 &= 0.87 \cdot 10^{-10} \, \rm{\frac{m}{s^2}} \,, \qquad & \frac{\sigma}{m} &= 0.01 \, \rm{\frac{cm^2}{g}} \,, \\
      r_0 &= 50\,\rm{kpc} \,, \qquad & \bar{a} &= 10^{-12}\, \rm{\frac{m}{s^2}} \,.
 \end{alignedat}
\end{equation}
In terms of $ \bar{\alpha} $, $ \Lambda $, $ m $, and $ \lambda_4 $, this gives:
\begin{equation}
 \begin{alignedat}{2}
       \bar{\alpha} &= 1.3 \cdot 10^{-6} \,, \qquad & m &= 6.4\,\rm{\mu eV} \,, \\
      \Lambda &= 0.5\,\rm{MeV} \,, \qquad & \lambda_4 &= 1.7\cdot10^{-20}\,.
 \end{alignedat}
\end{equation}
In particular, $ a_0 $ and $ \sigma/m $ are chosen to match the fiducial parameters from Ref.~\cite{Berezhiani2018}.
The characteristic length scale $ r_0 $ of $ \rho_{\rm{DM}-} $ is chosen to be on the order of the superfluid core sizes discussed in Ref.~\cite{Berezhiani2018}.
The minimum acceleration scale $ \bar{a} $, below which the {\sc MOND}ian {\sc EFT} breaks down, is chosen to be around the value where dwarf spheroidals start to deviate from {\sc MOND}ian behavior in Ref.~\cite{Lelli2017b}.
Thus, dwarf spheroidals may show non-{\sc MOND}ian behavior in our model.
This possibility was already mentioned in Ref.~\cite{Berezhiani2015} in relation to a different minimum acceleration scale that we discuss in Appendix~\ref{sec:origin}.
Of course, the existence of a minimum acceleration scale does not directly predict how dwarf spheroidals behave in {\sc SFDM}.
It merely gives an idea why those may not follow standard {\sc MOND}ian behavior.
Investigating this in more detail is beyond the scope of the present work.

\section{An explicit numerical example: The Milky Way}
\label{sec:pheno}

We now consider an explicit numerical example for our two-field {\sc SFDM} model.
We build on the Milky Way model from Ref.~\cite{Hossenfelder2020} in standard {\sc SFDM} that we already used in Sec.~\ref{sec:prob:mondlimit}.
For non-relativistic equilibrium solutions in galaxies, we can reuse the code from Ref.~\cite{Hossenfelder2020} with some modifications.
Namely, the $ \theta_+ $ equation of motion in two-field {\sc SFDM} is the same as the $ \theta $ equation in standard {\sc SFDM} if we set $\bar{\beta} = 0$.
Also, the Poisson equation for $ \phi_{\rm{N}} $ has the same form as in standard {\sc SFDM}, but we need to replace the dark matter energy density $ \rho_{\rm{DM}} $ from Eq.~\eqref{eq:rhoDMstandard} with the energy density $ \rho_{\rm{DM}-} $ from Eq.~\eqref{eq:rhoDM}.
The energy density $ \rho_{\rm{DM}+} $ is negligible for our choice of parameters, as we will see below.
There is also a technical modification we need to make:
In Ref.~\cite{Hossenfelder2020}, we rewrote the equation for $ \hat{\mu} $ in terms of $ \hat{\mu}_{\rm{tmp}} \equiv \hat{\mu} + \Delta \hat{\mu} $ with constant $ \Delta \hat{\mu} $ to avoid numerical issues with Mathematica.
In our model, we do not include the finite-temperature corrections associated with $ \bar{\beta} $.
As a result, $ \hat{\mu} $ enters the $ \theta_+ $ equation with opposite sign compared to how it enters the $ \theta $ equation in standard {\sc SFDM} and we need to switch the sign of $ \Delta \hat{\mu} $ to avoid the same numerical issue in Mathematica.

As boundary condition for $\hat{\mu}/m$, we impose $\mu_\infty/m = 1.25 \cdot 10^{-7}$ at $r_\infty = 100\,\rm{kpc}$.
For standard {\sc SFDM}, we use the same values as in Sec.~\ref{sec:prob:mondlimit}, namely $\mu_\infty/m = 1.25 \cdot 10^{-8}$ and $r_\infty = 100\,\rm{kpc}$.
For this choice, the dark matter densities of standard {\sc SFDM} and two-field {\sc SFDM} are comparable in magnitude, as we will see below.

\begin{figure}
 \centering
 \includegraphics[width=.49\textwidth]{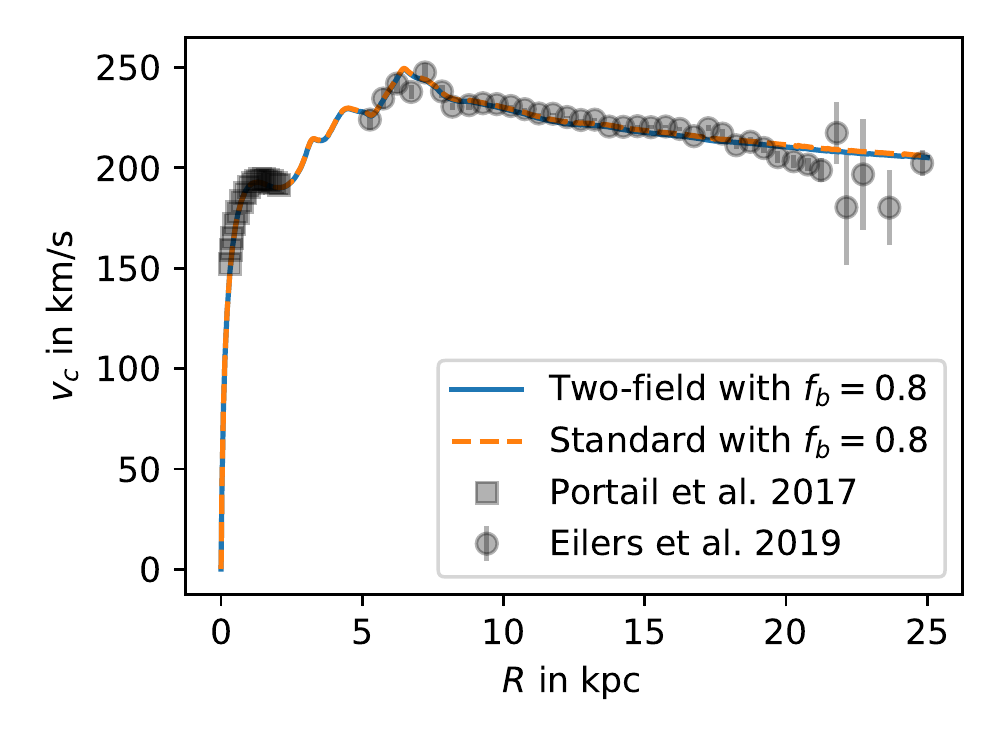}
 \includegraphics[width=.49\textwidth]{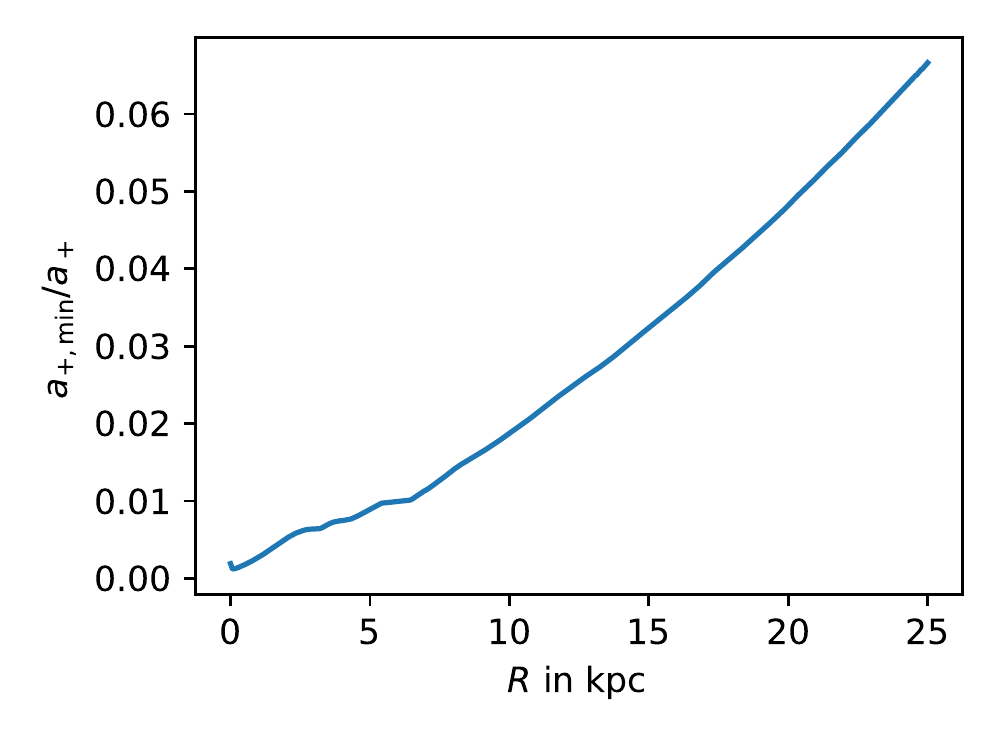}
 \caption{
     Left: Milky Way rotation curve of two-field {\sc SFDM} (solid blue line) and standard {\sc SFDM} (dashed orange line) for the Milky Way model from Ref.~\cite{Hossenfelder2020} with $f_b = 0.8$.
     Also shown is the rotation curve data from Ref.~\cite{Portail2017, Eilers2019} in black, both adjusted to match the assumptions of Ref.~\cite{McGaugh2019b}.
     Right: The quantity $ a_{+\rm{min}}/|\vec{a}_+| $ at $z=0$ that needs to be smaller than $ 1 $ for a stable equilibrium solution.
}
 \label{fig:SFDMrot}
\end{figure}

\begin{figure}
 \centering
 \includegraphics[width=.49\textwidth]{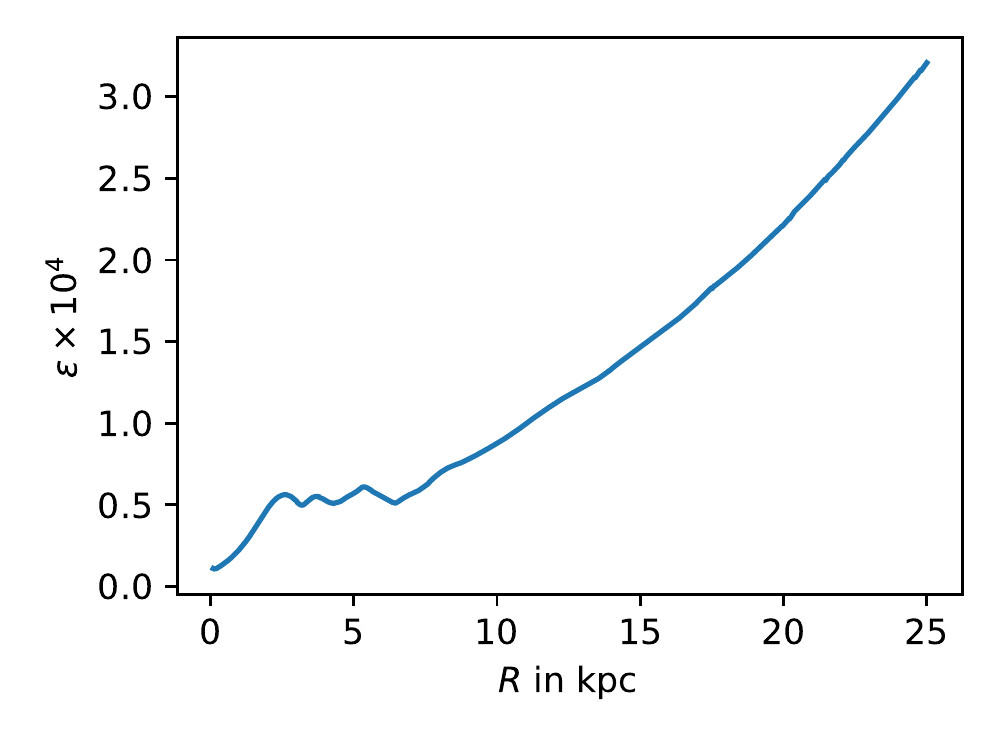}
 \includegraphics[width=.49\textwidth]{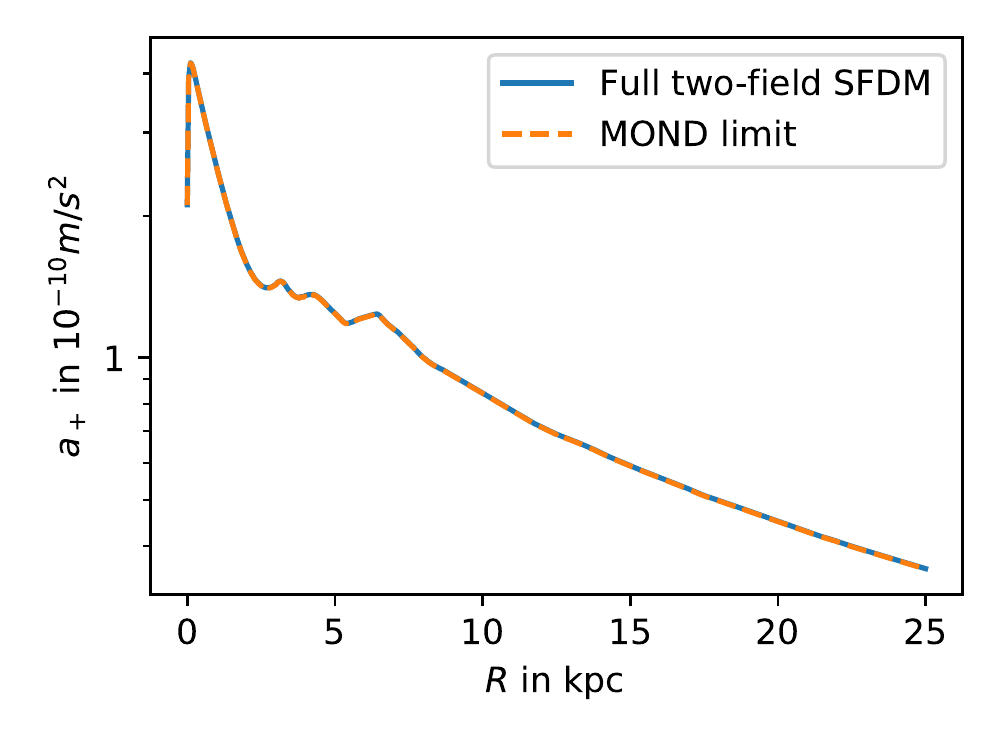}
 \caption{
    Left: The quantity $ \varepsilon = (2m\hat{\mu})/(\vec{\nabla} \theta_+)^2 $ that controls the idealized {\sc MOND} limit of two-field {\sc SFDM} at $z = 0$ for the Milky Way model from Ref.~\cite{Hossenfelder2020} with $ f_b = 0.8 $.
    Note the different scale of the vertical axis compared to Fig.~\ref{fig:MONDregime}, left.
     Right: The phonon acceleration $\vec{a}_+$ at $ z = 0 $ for full two-field {\sc SFDM} (solid blue line) and for the idealized {\sc MOND} limit from Eq.~\eqref{eq:MONDeq} (dashed orange line).
}
 \label{fig:MONDregimeImproved}
\end{figure}

\begin{figure}
 \centering
 \includegraphics[width=.49\textwidth]{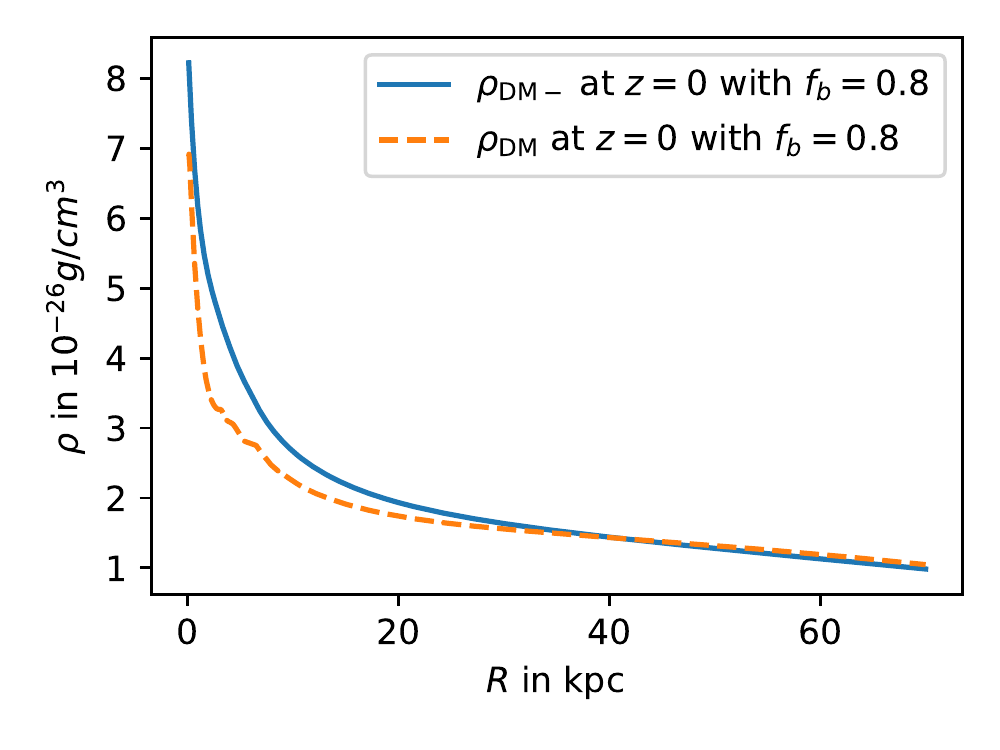}
 \includegraphics[width=.49\textwidth]{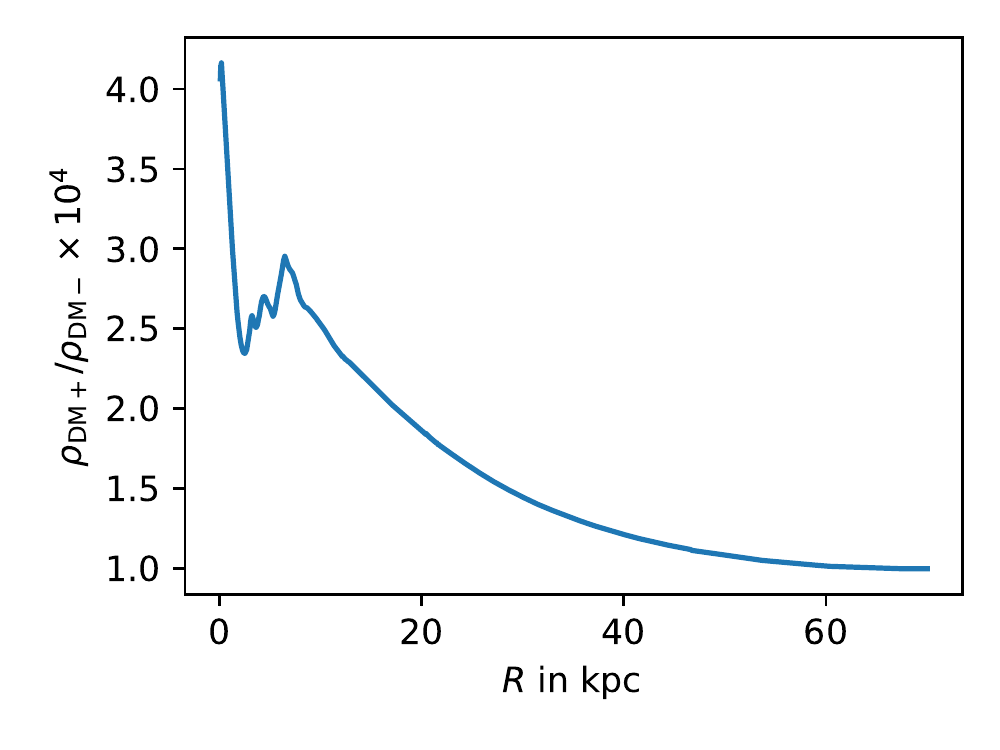}
 \caption{
     Left: The superfluid density of two-field {\sc SFDM} ($\rho_{\rm{DM}-}$, solid blue line) and of standard {\sc SFDM} ($\rho_{\rm{DM}}$, dashed orange line) at $z=0$ for the Milky Way model from Ref.~\cite{Hossenfelder2020} with $ f_b = 0.8 $.
     Right: The ratio $ \rho_{\rm{DM}+} / \rho_{\rm{DM}-} $ of the two contributions to the superfluid's energy density of two-field {\sc SFDM} at $ z = 0 $.
}
 \label{fig:rhoDM}
\end{figure}

In Fig.~\ref{fig:SFDMrot}, left, we compare the Milky Way rotation curves of two-field {\sc SFDM} and standard {\sc SFDM}.
Both are very close to each other for $ R < 25\,\rm{kpc} $.
Also, both have phonon accelerations close to the idealized {\sc MOND} limit as is shown in Fig.~\ref{fig:MONDregime}, right, for standard {\sc SFDM} and in Fig.~\ref{fig:MONDregimeImproved}, right, for two-field {\sc SFDM}.
For standard {\sc SFDM}, this is the result of a numerical coincidence (see Sec.~\ref{sec:prob:mondlimit} and Appendix~\ref{sec:mondregimecoincidence}).
Indeed, the quantity $ (2 m \hat{\mu})/(\vec{\nabla} \theta)^2 $ that needs to be small in the idealized {\sc MOND} limit is not small throughout the Milky Way, as shown in Fig.~\ref{fig:MONDregime}, left.
In contrast, the analogous quantity in two-field {\sc SFDM}, $ (2 m \hat{\mu})/(\vec{\nabla} \theta_+)^2 $, is small throughout the Milky Way, as shown in Fig.~\ref{fig:MONDregimeImproved}, left.
Thus, two-field {\sc SFDM} does have a proper {\sc MOND} limit, i.e. it has a {\sc MOND}ian phonon acceleration as well as a {\sc MOND}ian {\sc EFE} (see Sec.~\ref{sec:prob:mondlimit}).

In Sec.~\ref{sec:model}, we saw that the equilibrium solutions on galactic scales are stable only if the acceleration $ \vec{a}_+ $ due to $ \theta_+ $ is larger than $ a_{+\rm{min}} = a_0 (\lambda_4/\bar{\alpha}^2) (m/\hat{\mu}) $.
Fig.~\ref{fig:SFDMrot}, right, shows that this condition is fulfilled in the inner parts of the Milky Way.
This is expected as the Newtonian baryonic acceleration $ a_b $ is around $ a_0 \approx 10^{-10} \, \rm{m}/\rm{s}^2 $ there and, by construction, the stability condition should fail only at $ a_b \lesssim \bar{a} = 10^{-12} \, \rm{m}/\rm{s}^2 $ (see Sec.~\ref{sec:params}).

In Fig.~\ref{fig:rhoDM}, right, we see that $ \rho_{\rm{DM}+} $ is negligible compared to $ \rho_{\rm{DM}-} $.
This is related to our model having a proper {\sc MOND} limit, as discussed in Sec.~\ref{sec:model}.
A proper {\sc MOND} limit requires small $ m^2/\bar{\alpha} $ which in turn makes $ \rho_{\rm{DM}+} $ small.
Indeed, $ m^2/\bar{\alpha} $ is much smaller in our case than in standard {\sc SFDM}.
Concretely, $ m^2/\bar{\alpha} \approx 10^{-5}\,\rm{eV}^2 $ in our case and  $ m^2/\bar{\alpha} \approx 10^{-1}\,\rm{eV}^2 $ in standard {\sc SFDM} with the fiducial parameters from Ref.~\cite{Berezhiani2018}.

In Fig.~\ref{fig:rhoDM}, left, we compare the dark matter density of two-field {\sc SFDM} and standard {\sc SFDM}.
Both are comparable in magnitude, but standard {\sc SFDM} gives a more cored profile.
As already mentioned in Sec.~\ref{sec:halo}, this is a consequence of our choice of the $ |\phi_-|^4 $ interaction term, which gives $ \rho_{\rm{DM}} \propto \hat{\mu}/m $.
Other choices are certainly possible, e.g. a $ |\phi_-|^6 $ interaction gives $ \rho_{\rm{DM}-} \propto \sqrt{\hat{\mu}/m} $ which is closer to the energy density in standard {\sc SFDM} \cite{Berezhiani2015}.
The difference between different density profiles is not important for the Milky Way rotation curve at $ R < 25\,\rm{kpc} $, since the gravitational pull due to this energy density is subdominant there \cite{Hossenfelder2020}.
For observables at larger radii, like strong lensing, this difference may be more important.
At these radii, the superfluid's gravitational pull is dominant and the energy density of our model may be very different from standard {\sc SFDM} due to its different shape -- even if both are comparable at smaller radii.
If needed, we can adjust the boundary condition $\mu_\infty$ of $\hat{\mu}$, or try a different interaction term like the $|\phi_-|^6$ term mentioned above.

\section{Size of the superfluid core}
\label{sec:RTRNFW}

A central idea in {\sc SFDM} is that the superfluid at the center of a galaxy breaks down at larger radii.
Beyond the superfluid core, the dark matter profile is assumed to be that of a standard {\sc NFW} halo \cite{Berezhiani2018}.
Ref.~\cite{Berezhiani2018} gives two different ways to estimate the size of the superfluid core for spherically symmetric systems -- the ``{\sc NFW} radius'' and the ``thermal radius''.
In Ref.~\cite{Hossenfelder2020}, this was extended to axisymmetric systems.
In this section, we aim to extend these estimates to our two-field {\sc SFDM} model.
We find that this seems to work reasonably well for the {\sc NFW} radius.
However, considering also the thermal radius raises questions about the interpretation of the {\sc NFW} and thermal radii and, more generally, about how to correctly estimate the superfluid core's size.

\subsection{The {\sc NFW} radius}

The {\sc NFW} radius $ R_{\rm{NFW}} $ is defined as the radius where the density and pressure of the superfluid equal the density and pressure of a standard {\sc NFW} halo \cite{Berezhiani2018}.
In axisymmetric systems, there are in principle different {\sc NFW} radii in $ R $ direction and in $ z $ direction.
However, those different radii are typically very close so that the difference does not matter for practical purposes \cite{Hossenfelder2020}.

We can straightforwardly adapt this estimate for the superfluid core's size to two-field {\sc SFDM}.
For the superfluid density we can use $ \rho_{\rm{DM}-} = (2 m^3 \hat{\mu})/\lambda_4 = 2 (M_{\rm{Pl}}/r_0)^2 (\hat{\mu}/m) $ and for the superfluid pressure we can use $ P_{\rm{DM}-} = (m \hat{\mu})^2/\lambda_4 = (M_{\rm{Pl}}/r_0)^2 (\hat{\mu}/m)^2 $ (see e.g. Ref.~\cite{Sharma2019}).
The formulas for the pressure and density of the {\sc NFW} halo can be taken directly from Ref.~\cite{Berezhiani2018}.
Then, we can use the same procedure as in Refs.~\cite{Berezhiani2018, Hossenfelder2020}.
As a concrete example, consider the Milky Way model from Sec.~\ref{sec:pheno}.
We find for the {\sc NFW} radius in $ R $ direction:
\begin{align}
 R_{\rm{NFW}} = 73\,\rm{kpc} \,.
\end{align}
With this, the total dark matter mass inside the virial radius is $ M_{\rm{DM}}^{200} = 1.3 \cdot 10^{12}\,M_\odot $.
These values are similar to those obtained in standard {\sc SFDM} with the boundary conditions used in Sec.~\ref{sec:prob:mondlimit}, namely $ R_{\rm{NFW}} = 66\,\rm{kpc} $ and $ 1.2\cdot10^{12}\,M_\odot $ \cite{Hossenfelder2020}.
We expect these values to depend strongly on the choice of the boundary condition $\mu_\infty/m$.
This is in contrast to observables at smaller radii, like the rotation curve at $R < 25\,\rm{kpc}$, which depend only very weakly on $\mu_\infty/m$.
This was demonstrated in Ref.~\cite{Hossenfelder2020} for standard {\sc SFDM} and we expect the same to be true for two-field {\sc SFDM}.

Here, we assumed that the term $f(K_+ + K_- -m^2)$ from our Lagrangian does not contribute to the energy density and pressure.
For the energy density, we saw in Sec.~\ref{sec:pheno} that this is justified.
However, there is a significant contribution to the radial pressure from this term.
Concretely, in the {\sc MOND} limit $ \varepsilon \ll 1 $, spatial gradients of $ \theta_+ $ give the radial pressure,
\begin{align}
 \label{eq:PSF+}
 \frac43 \Lambda |\vec{\nabla} \theta_+|^3 = \frac43 \frac{M_{\rm{Pl}}^2 a_+^3}{a_0} \,.
\end{align}
It is not quite clear whether or not this pressure should be included when matching the superfluid pressure to the {\sc NFW} pressure.
On the one hand, our superfluid corresponds to the $ \phi_- $ field, not $ \theta_+ $, so maybe we should not include this pressure due to $ \theta_+ $.
On the other hand, $ \theta_- $ and $ \theta_+ $ mix in the $ f(K_+ + K_- - m^2) $ term, so this distinction is not sharp.
More generally, we may anyway need more matching conditions than standard {\sc SFDM} since we have more fields.
However, numerically, including the pressure term from Eq.~\eqref{eq:PSF+} does not significantly change the {\sc NFW} radius and the total dark matter mass.
The {\sc NFW} radius becomes $ 78\,\rm{kpc} $ and the total dark matter mass becomes $ 1.4\cdot10^{12}\,M_\odot $.
Thus, this point does not make a big difference for the basic phenomenology.

\subsection{The thermal radius}
\label{sec:RT}

The second estimate of the superfluid core's size from Ref.~\cite{Berezhiani2018} is the thermal radius $ R_T $.
This radius is defined as the radius where the local self-interaction rate $ \Gamma $ drops below the inverse dynamical time $ t_{\rm{dyn}}^{-1} $.
The idea is that, to sustain equilibrium, the timescale of the local self-interaction rate needs to be larger than the timescale of the perturbations, which is assumed to be $ t_{\rm{dyn}} $.
Here, $ \Gamma = (\sigma / m) \, \mathcal{N} \, v \, \rho $, where $ \sigma $ is the self-interaction cross-section, $ \mathcal{N} = (\rho/m) (2\pi/mv)^3 $
is the Bose enhancement factor, $ v $ is the average velocity of the particles, and $ \rho $ is the dark matter density.

In spherically symmetric systems, Ref.~\cite{Berezhiani2018} takes $ t_{\rm{dyn}} = r/v $ with the spherical radius $ r $.
This gives
\begin{align}
 \label{eq:RTcond}
 (2 \pi)^3 \, \frac{(\sigma/m)}{m^4} \frac{\rho^2}{v^2} = \frac{v}{R_T} \,.
\end{align}
The quantities $ v $, $ \rho $ and $ \sigma/m $ are numerically similar in two-field {\sc SFDM} and in standard {\sc SFDM} for the parameter values adopted here.
The reason is that we chose $ \sigma/m $ to match the fiducial numerical value from Ref.~\cite{Berezhiani2018}, and the density and velocity can also be very similar, as we saw in Sec.~\ref{sec:pheno}.
However, $ m $ is very different.
The fiducial value from Ref.~\cite{Berezhiani2018} is $ m = 1\,\rm{eV} $ while our parameter choice gives the much smaller value $ m = 6.4\cdot10^{-6}\,\rm{eV} $.
Thus, at a given radius, the left-hand side of Eq.~\eqref{eq:RTcond} is about $ 10^{21} $ times larger than in standard {\sc SFDM} and there is probably no solution to Eq.~\eqref{eq:RTcond} on galactic scales.
I.e. we always have $ \Gamma \gg t_{\rm{dyn}}^{-1} $ on galactic scales.
At face value, this implies that there is no transition to a {\sc NFW} halo on galactic scales, in contrast to the standard expectation in {\sc SFDM} and in contrast to what the {\sc NFW} radius suggests.
This could mean that the superfluid phase extends to cosmological scales with a matching between the cosmological and the galactic solution at an intermediate scale.
Alternatively, the superfluid could end at a finite radius where density and pressure reach zero when $\hat{\mu}$ reaches zero (at $r = \mathcal{O}(r_0)$, see Sec.~\ref{sec:halo}).
In this case, the superfluid would resemble a giant non-relativistic boson star \cite{Tkachev1986, Colpi1986, Lee1996, Sharma2008}\footnote{
    Strictly speaking, $\hat{\mu}$ does not reach zero, i.e. boson stars don't have a surface \cite{Colpi1986}:
    Before $\hat{\mu}$ reaches zero, derivatives of $\rho_-$ become important, but we neglected those above.
    Our approximation breaks down when $\rho_{\rm{DM}-}$ drops to about $(m^3/\lambda_4 M_{\rm{Pl}})^2$, which, in our case, is much smaller than the cosmological dark matter density.
    Thus, for practical purposes $\hat{\mu}$ does reach zero on galactic scales.
}.

In Sec.~\ref{sec:params}, we chose $ \sigma/m $ in our model to be the same as in standard {\sc SFDM} with the fiducial parameters from Ref.~\cite{Berezhiani2018}.
Alternatively, we could have chosen $ (\sigma/m)/m^4 $ to be the same as in standard {\sc SFDM}.
In this case, our $ R_T $ would be closer to that of standard {\sc SFDM} and the discrepancy between the {\sc NFW} radius and the thermal radius in our model would be much smaller.
Still, as our actual choice of parameters illustrates, the thermal radius and the {\sc NFW} radius need not be close to each other in general.
Indeed, the {\sc NFW} radius can be calculated from only $r_0$ and $\hat{\mu}/m$, while the thermal radius also depends on $(\sigma/m)/m^4 \propto (\sigma/m)^{1/5} r_0^{16/5}$.

The above raises the question whether the thermal radius, the {\sc NFW} radius, or a completely different radius gives the physical size of a galaxy's superfluid core.
In standard {\sc SFDM} with the fiducial numerical parameters from Ref.~\cite{Berezhiani2018}, the thermal and {\sc NFW} radii are numerically reasonably close \cite{Berezhiani2018}.
Thus, for practical purposes, the difference may not be too important in standard {\sc SFDM}.
However, the thermal radius and the {\sc NFW} radius can differ significantly also in standard {\sc SFDM}, if we depart from the fiducial parameters from Ref.~\cite{Berezhiani2018}.\footnote{
    For example, assuming the {\sc MOND} limit $\varepsilon \ll 1$, the {\sc NFW} radius depends only on $m^2/\bar{\alpha}$, $a_0$, and $|\vec{a}_b|$, while the thermal radius also depends on $\sigma/m$.
}
Investigating this is left for future work.

\section{Discussion}
\label{sec:discussion}

As discussed in Sec.~\ref{sec:sfdm}, the field $ \theta $ of standard {\sc SFDM} serves a double role.
One role is to mediate a {\sc MOND}-like force, the other role is to provide the dark matter density.
In our two-field {\sc SFDM} model, these two roles are split between $ \theta_+ $ and $ \phi_- \propto \rho_- e^{-i \theta_-} $.
For $ \phi_- $ we take a standard superfluid Lagrangian with quartic interactions.
For $ \theta_+ $ we take the original Lagrangian from standard {\sc SFDM} but replace $ K_\theta \to K_+ + K_- $.
An alternative would have been to get rid of the standard {\sc SFDM} Lagrangian and instead take a standard {\sc MOND}ian Lagrangian for $ \theta_+ $, e.g. a {\sc RAQUAL}-type Lagrangian \cite{Bekenstein1984, Famaey2012, Bruneton2007b}
\begin{align}
 \mathcal{L} = \Lambda \sqrt{|K_+|} K_+ - \lambda \, \theta_+ \, \rho_b \,.
\end{align}
That is, a Lagrangian where the mass $ m $ does not occur and where $ \theta_+ $ and $ \theta_- $ are completely independent of each other, unlike in the $ f(K_+ + K_- -m^2) $ term in $ \mathcal{L}_{\rm{imp}} $.
Such a model could give a qualitatively similar phenomenology on galactic scales:
The field $ \theta_+ $ mediates a {\sc MOND}ian force and $ \phi_- $ provides the dark matter density.
Indeed, a model along these lines was proposed in Ref.~\cite{Khoury2015}.
This model has qualitative similarities to two-field {\sc SFDM}, but is also different in many ways.
For example, the field in Ref.~\cite{Khoury2015} that is analogous to our $ \phi_- $ field does not cluster on galactic scales, it only provides the average dark matter density of the cosmological background.
Also, to get enough lensing, matter couples to a metric that is disformally related to the Einstein metric in Ref.~\cite{Khoury2015}, while the {\sc SFDM} coupling $ \lambda \rho_b \theta_+ $ is closer to an effective metric that is conformally related to the Einstein metric.
Another difference is that, in our model, the $\phi_-$ field (responsible for the dark matter energy density) induces the {\sc MOND}ian regime of the $ \theta_+ $ field (responsible for the additional force on the baryons).
Indeed, the additional force due to $ \theta_+ $ is {\sc MOND}ian only inside the superfluid.
The reason is that the equation of motion of $ \theta_+ $ is a {\sc MOND}ian equation only if $ K_+ + K_- - m^2 \approx -(\vec{\nabla} \theta_+)^2 $ (see Sec.~\ref{sec:model}), which is possible only if the superfluid's chemical potential $ \mu $ cancels the $-m^2 $ term through the equilibrium value of $K_-$, namely $K_- \approx g^{00} \mu^2$.
In contrast, the field responsible for the additional force on the baryons in Ref.~\cite{Khoury2015} satisfies an equation of motion that is independent of the field providing the dark matter energy density.
There are likely also differences at larger scales, e.g. on cluster and cosmological scales, but we have not yet explored these scales in any detail in our model.

The standard {\sc SFDM} model from Ref.~\cite{Berezhiani2015} can be seen as the successor of the two-field model from Ref.~\cite{Khoury2015} with the aim of finding a common origin for the {\sc MOND}ian force and the dark matter energy density.
Our discussion in Sec.~\ref{sec:sfdm} now suggests that it is tough to find such a common origin in the form of a single field.
Our analysis indicates that it might be of advantage to switch back to a model with two fields, but with phenomenology closer to standard {\sc SFDM}.

Our model is also related to the two-field model proposed in Sec.~7 of Ref.~\cite{Mistele2019}.
This model from Ref.~\cite{Mistele2019} contains fields $ \theta_+ $ and $ \theta_- $ similar to our fields $ \theta_+ $ and $ \theta_- $ in that one of the fields has an exact shift symmetry $ \theta_- \to \theta_- + \rm{const}. $, while the shift symmetry of the other field is broken by the baryon coupling.
However, this model from Ref.~\cite{Mistele2019} solves only the equilibrium problem (see Sec.~\ref{sec:prob:equilibrium}), but not the {\sc MOND} limit problem (see Sec.~\ref{sec:prob:mondlimit}) or the stability problem (see Sec.~\ref{sec:prob:stability}).

Finally, let us comment on the interpretation of our two-field model as a superfluid model.
Our Lagrangian is the sum of a standard self-interacting superfluid Lagrangian for $\phi_-$ and the non-standard $f(K_+ + K_- -m^2) - \lambda \, \rho_b \, \theta_+$ part.
The first part simply gives a superfluid with much of the standard superfluid phenomenology.
For example, it allows for frictionless movement of subsonic perturbers.
The second part is responsible for the {\sc MOND}ian force carried by $\theta_+$.
In this second part, the role of $\theta_-$ is mainly to induce the {\sc MOND} regime for $\theta_+$ when the superfluid is in the condensed phase.
That is, in the superfluid phase, the chemical potential cancels the $m^2$ in $f(K_+ + K_- -m^2)$, as mentioned above.
However, in equilibrium, the superfluid itself is not affected by this non-standard part.
Also, this second part does not significantly contribute to the total energy density.
Therefore, we interpret this non-standard part as responsible for a {\sc MOND}ian force, not as part of the superfluid.
This is different from standard {\sc SFDM} where the analogous part, $f(K_\theta - m^2) - \lambda \, \rho_b \, \theta$, is responsible for both the superfluid and the {\sc MOND}ian force.

For standard {\sc SFDM}, Ref.~\cite{Berezhiani2015} has shown how the $f(K_\theta - m^2)$ term can be obtained from an unconventional superfluid-like Lagrangian of a complex field $\phi \propto \rho \, e^{-i \theta}$.
This works by giving $\phi$ a non-canonical kinetic term and then integrating out its modulus $\rho$.
Similarly, the $f(K_+ + K_- - m^2)$ term in our model can be obtained from a Lagrangian of two complex fields $\phi_-$ and $\phi_+$ by integrating out the modulus of $\phi_+ \propto \rho_+ \, e^{-i \theta_+}$.
This is discussed in Appendix~\ref{sec:origin}.
In contrast, as in standard {\sc SFDM}, the origin of the symmetry-breaking coupling $\lambda \, \rho_b \, \theta_+$ is unclear.

\section{Conclusion}
\label{sec:conclusion}

In this paper, we have shown that the double role of the phonon field $\theta$ in standard {\sc SFDM} as a carrier of both the {\sc MOND}ian force and the superfluid's energy density leads to problems.
Concretely, this double role is in tension with having a proper {\sc MOND} limit and with having an equilibrium that is valid much longer than the dynamical timescale of galaxies.
We have proposed an improved model that avoids these problems and also addresses a well-known instability issue in standard {\sc SFDM} in a less ad-hoc way.
This model works by splitting the two roles of the phonon field $ \theta $ between two different fields.
One of these carries the {\sc MOND}ian force, the other carries the superfluid's energy density.

Using the Milky Way as an explicit example, we have demonstrated that our model's phenomenology on galactic scales can be close to that of standard {\sc SFDM} while avoiding the tensions due to the phonon field's double role in standard {\sc SFDM}.
However, we found that the transition from the superfluid phase at smaller radii to the normal phase at larger radii needs to be revisited.
The reason is that, in general, the two standard estimates for the radius of this transition, the {\sc NFW} radius and the thermal radius, can differ wildly.
In principle, this issue is already present in standard {\sc SFDM}, although it may not be too important in practice for the fiducial numerical parameters from Ref.~\cite{Berezhiani2018}.
Future work should clear up the interpretation of these different estimates and identify how to correctly estimate the size of the superfluid core in {\sc SFDM}.

\section*{Acknowledgements}
\label{sec:acknowledgements}

I thank Sabine Hossenfelder for valuable discussions and for reading the manuscript.
I am grateful for financial support from FIAS.

\begin{appendices}

\section{Validity of the {\sc MOND} regime in standard {\sc SFDM}}
\label{sec:mondregimecoincidence}

In standard {\sc SFDM}, the quantity $\varepsilon = ( 2 m \hat{\mu})/(\vec{\nabla} \theta)^2$ controls the idealized {\sc MOND} limit, i.e. whether or not the $\theta$ equation of motion has a standard {\sc MOND}ian form.
Thus, we would expect that the acceleration $\vec{a}_\theta$ due to $\theta$ has a {\sc MOND}ian form only if $\varepsilon \ll 1$.
However, we saw in an explicit example in Sec.~\ref{sec:prob:mondlimit} that $\vec{a}_\theta$ can have a {\sc MOND}ian form even if $\varepsilon$ is larger than 1.
Here, we investigate why $ \vec{a}_\theta $ computed in the full {\sc SFDM} model is numerically close to $ \vec{a}_\theta $ computed from the idealized {\sc MOND} limit Eq.~\eqref{eq:MONDeq} despite $\varepsilon$ being larger than 1.

To this end, we employ the so-called no-curl-approximation.
This means the following.
The equation for $ \theta $ is of the form $ \vec{\nabla} ( g \vec{a}_\theta ) = \vec{\nabla} \vec{a}_b $ with some function $ g $.
Thus, we have $ g \vec{a}_\theta = \vec{a}_b $ up to a term that is the curl of a vector field.
Neglecting this curl term is the no-curl-approximation.
This approximation was shown to be a good approximation in {\sc SFDM} in Ref.~\cite{Hossenfelder2020}.
The advantage of this approximation is that we can algebraically solve for $\vec{\nabla} \theta$ in terms of $ \hat{\mu} $ and $ \vec{a}_b $.

\begin{figure}
 \centering
 \includegraphics[width=.49\textwidth]{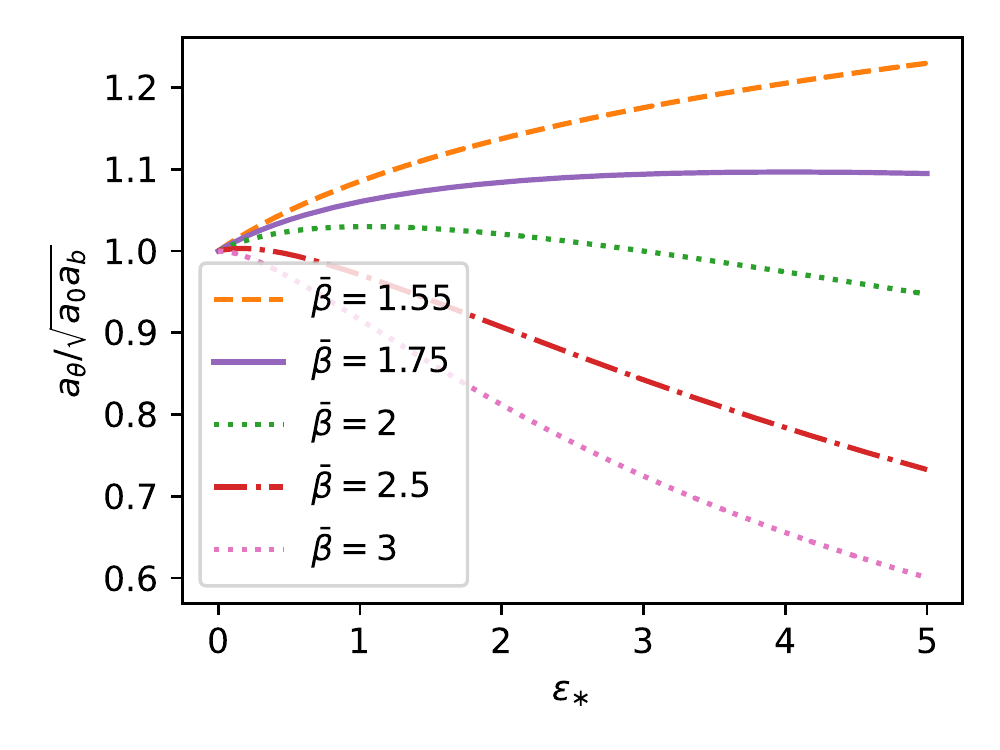}
 \includegraphics[width=.49\textwidth]{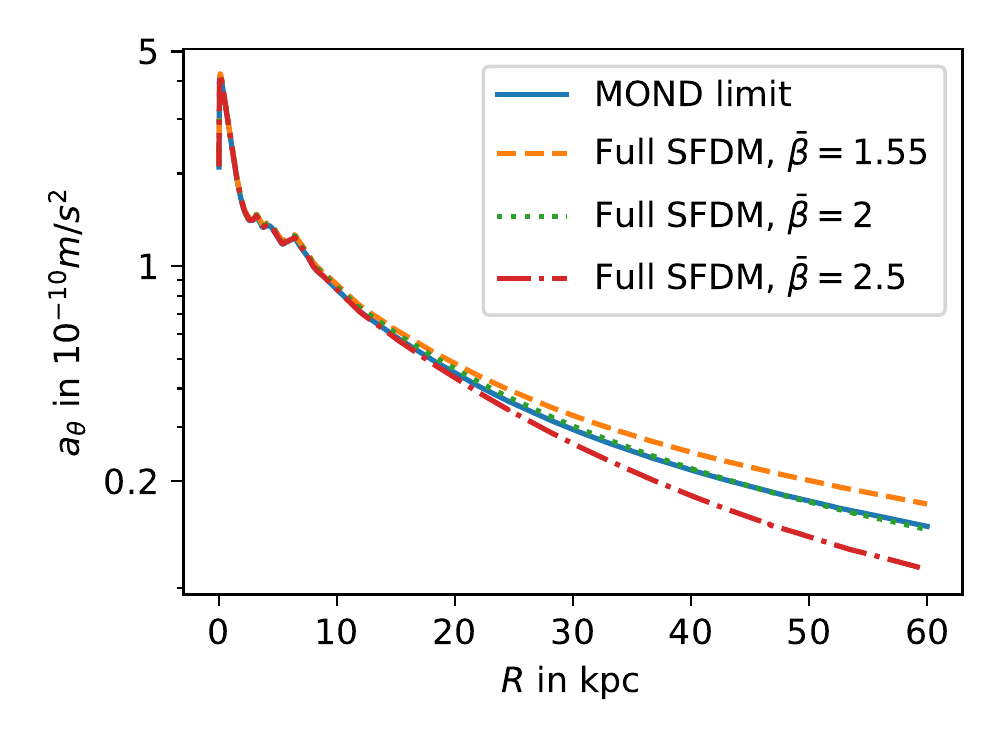}
 \caption{
     Left: The quantity $\sqrt{x} = |\vec{a}_\theta|/\sqrt{a_0 |\vec{a}_b|}$ in the no-curl-approximation of standard {\sc SFDM} as a function of $\varepsilon_* = (2 m \hat{\mu})/(\bar{\alpha} M_{\rm{Pl}} |\vec{a}_b|)$.
     For $\sqrt{x} = 1$, the phonon force has a {\sc MOND}ian form.
     Right: Phonon accelerations $\vec{a}_\theta$ at $z=0$ for the Milky Way model from Ref.~\cite{Hossenfelder2020} in the full standard {\sc SFDM} model with $ \bar{\beta} = 2.5 $ (dash-dotted red line), $\bar{\beta} = 1.55$ (dashed orange line), the fiducial value $ \bar{\beta} = 2 $ from Ref.~\cite{Berezhiani2018} (dotted green line), and for the idealized {\sc MOND} limit (solid blue line).
     The parameters are as in Sec.~\ref{sec:prob:mondlimit}, i.e. $f_b=0.8$, $\mu_\infty/m=1.25\cdot10^{-8}$, and $r_\infty = 100\,\rm{kpc}$.
}
 \label{fig:MONDregimeCoincidence}
\end{figure}

Concretely, in the no-curl-approximation, we have
\begin{align}
  |\vec{a}_\theta| = \sqrt{a_0 |\vec{a}_b|} \cdot \sqrt{x} \,,
\end{align}
where $x = x(\varepsilon_*, \bar{\beta})$ is determined by the cubic equation
\begin{align}
 \label{eq:xeq}
  0 &= x^3 + 2 \left(\frac{2\bar{\beta}}{3} - 1\right) \varepsilon_* \cdot x^2 + \left( \left(\frac{2\bar{\beta}}{3}-1\right)^2 (\varepsilon_*)^2 - 1\right) x - \left(\bar{\beta}-1\right) \varepsilon_* \,,
\end{align}
with
\begin{align}
  \varepsilon_* &\equiv \frac{2 m \hat{\mu}}{\bar{\alpha} M_{\rm{Pl}} |\vec{a}_b|} \,.
\end{align}
The phonon acceleration is approximately {\sc MOND}ian whenever $\sqrt{x}$ is approximately 1.
Whether or not this is the case is determined by $\bar{\beta}$ and $\varepsilon_*$.

As discussed in Sec.~\ref{sec:prob:mondlimit}, we expect $ \vec{a}_\theta$ to have a {\sc MOND}ian form if $\varepsilon = (2 m \hat{\mu})/(\vec{\nabla} \theta)^2$ is small.
Indeed, Eq.~\eqref{eq:xeq} has the solution $x = 1$ in this limit.
The reason is that, for $\varepsilon \ll 1$, we have $ (\vec{\nabla} \theta)^2 \approx \bar{\alpha} M_{\rm{Pl}} |\vec{a}_b|$ so that $\varepsilon_* \approx \varepsilon$.
Thus, in this case $\varepsilon_*$ is small and Eq.~\eqref{eq:xeq} reduces to $x^3 - x = 0$.\footnote{
    The $x=0$ solution of this equation is spurious.
    The reason is that Eq.~\eqref{eq:xeq} was obtained by multiplying the $\theta$ equation of motion by a quantity that is proportional to $x$ in the limit of small $\varepsilon_*$.
}
This is the standard {\sc MOND} limit of {\sc SFDM} for $\varepsilon \ll1 $.

However, we saw in Fig.~\ref{fig:MONDregime} that $\vec{a}_\theta$ can have a {\sc MOND}ian form even if $\varepsilon$ is not small.
To understand this, we now turn to more general solutions of Eq.~\eqref{eq:xeq}.
This equation can be solved analytically, but the resulting expression is not very illuminating so we do not show it here.
Instead, we show solutions $\sqrt{x} = \sqrt{x}(\varepsilon_*, \bar{\beta})$ determined by Eq.~\eqref{eq:xeq} in Fig.~\ref{fig:MONDregimeCoincidence}, left.
The horizontal axis represents different values of $\varepsilon_*$ and the different lines represent different values of $\bar{\beta}$.
Usually, $\bar{\beta}$ is taken to be the same across different galaxies \cite{Berezhiani2015, Berezhiani2018}.
In contrast, $\varepsilon_*$ is constant neither within a galaxy (since it depends on $\vec{x}$ through $\hat{\mu}(\vec{x})$ and $\vec{a}_b(\vec{x})$) nor across different galaxies (since $\hat{\mu}$ and $\vec{a}_b$ are different in different galaxies).
With the fiducial parameters from Ref.~\cite{Berezhiani2018}, we have (see also Eq.~\eqref{eq:epsilon})
\begin{align}
 \varepsilon_* = \frac{2 m^2}{\bar{\alpha}} \frac{10^{-6}}{a_0 M_{\rm{Pl}}} \left(\frac{a_0}{10 |\vec{a}_b|}\right) \left(10^7 \frac{\hat{\mu}}{m}\right) \approx 0.8 \cdot \left(\frac{a_0}{10 |\vec{a}_b|}\right) \left(10^7 \frac{\hat{\mu}}{m}\right)\,.
\end{align}
In galaxies, a typical value of $\hat{\mu}/m$ is $10^{-7}$ and the phonon force becomes important when $|\vec{a}_b|$ is smaller than $a_0$ so that $\varepsilon_*$ is roughly of order 1.
In Fig.~\ref{fig:MONDregimeCoincidence}, left, we show values of $\varepsilon_*$ between 0 and 5.
Within a galaxy, different values of $\varepsilon_*$ correspond to different radii.
As a rough estimate, for $|\vec{a}_b| \propto 1/r^2$ and $\hat{\mu}/m \propto 1/r$, we have
\begin{align}
\varepsilon_* \propto r \,.
\end{align}

Consider now $\bar{\beta} = 2$, i.e. the fiducial value from Ref.~\cite{Berezhiani2018}.
From Fig.~\ref{fig:MONDregimeCoincidence}, left, we see that $\sqrt{x}$ is always close to 1 for $\varepsilon_*$ between 0 and 5.
In contrast, for $\bar{\beta}$ larger or smaller than 2, $\sqrt{x}$ is close to 1 only for $\varepsilon_* \ll 1$, i.e. in the standard {\sc MOND} limit.
That is, for $\bar{\beta} = 2$, the phonon force can be {\sc MOND}ian across a relatively wide range of galaxies and radii -- even if $\varepsilon_*$ is not small.
For other values of $\bar{\beta}$, deviations from a {\sc MOND}ian force are much larger at $\varepsilon_* \gtrsim 1$.
For example, if we choose $\bar{\beta} = 2.5$ or $\bar{\beta} = 1.55$, the phonon force $\vec{a}_\theta$ deviates from a {\sc MOND}ian force already at significantly smaller radii compared to the $\bar{\beta} = 2 $ case.
This is illustrated in Fig.~\ref{fig:MONDregimeCoincidence}, right.

Thus, even if $\varepsilon$ is not small, {\sc SFDM} can have a {\sc MOND}ian phonon force.
But only if $\bar{\beta}$ is close to 2 and (since $x \to 0$ for $\varepsilon_* \to \infty$) only if $\varepsilon_*$ is not much larger than 1.
For the Milky Way model discussed in Sec.~\ref{sec:prob:mondlimit} with the fiducial numerical parameters from Ref.~\cite{Berezhiani2018}, these conditions are fulfilled at $R \lesssim 25\,\rm{kpc}$ so that the phonon force is approximately {\sc MOND}ian.
However, this is a numerical coincidence and not by construction of the model.
Through $\varepsilon_*$ and $\bar{\beta}$, this numerical coincidence depends on the model parameters $\bar{\beta}$, $\bar{\alpha}$, and $m$, on the shape and size of the baryonic density, and on the boundary condition $\mu_\infty$.
Thus, this numerical coincidence will not work if, for example, $m^2/\bar{\alpha}$ is too large, if $\mu_\infty$ is too large, or, as already discussed, if $\bar{\beta}$ is not close to 2.

\section{Origin of $ f $}
\label{sec:origin}

Above, we have taken the form of the function $ f $ as given -- both in the standard {\sc SFDM} Lagrangian from Eq.~\eqref{eq:L} and in the two-field {\sc SFDM} Lagrangian from Eq.~\eqref{eq:Limp}.
However, due to the square root in $f$, both Lagrangians are ill-defined whenever their argument is zero, i.e. whenever $ K_+ + K_- - m^2 = 0 $ (in two-field {\sc SFDM}) or $ K_\theta - m^2 = 0 $ (in standard {\sc SFDM} without finite-temperature corrections).
In the inner parts of galaxies, the argument of $f$ is always negative so that we may ignore this problem.
But in the outer parts of galaxies or in cosmology, this argument may become positive.
To address this, Ref.~\cite{Berezhiani2015} proposed a Lagrangian which does not suffer from this problem.
It reads
\begin{align}
 \label{eq:Lorigstandard}
 \mathcal{L}_{\rm{orig}} = \frac12 \bar{K} + \frac{\Lambda^4}{6 (\Lambda_c^2 + \rho^2)^6} \bar{K}^3 - \lambda \, \theta \, \rho_b \,,
\end{align}
where $\rho$ is a new field, $ \Lambda_c $ is a constant, and
\begin{subequations}
\begin{align}
 \bar{K} &= K_\rho + \rho^2 (K_\theta - m^2) \,, \\
  K_\rho &= \nabla^\alpha \rho \nabla_\alpha \rho \,.
\end{align}
\end{subequations}
On galactic scales, the $\rho$ equation of motion has a solution $ \rho^2 = \Lambda \sqrt{|K_\theta - m^2|} $, if we can neglect derivatives of $ \rho $ and if $ \rho^2 \gg \Lambda_c^2 $.
With this solution, we recover the Lagrangian from Eq.~\eqref{eq:L} with the $ f(K_\theta - m^2) $ term as an effective Lagrangian.
The condition $ \rho^2 \gg \Lambda_c^2 $ gives a minimum acceleration below which this Lagrangian is invalid.
Namely, in the {\sc MOND} limit $ (\vec{\nabla} \theta)^2 \gg 2 m \hat{\mu} $, the effective Lagrangian is valid only if,\footnote{This equation appears in Ref.~\cite{Berezhiani2015} as Eq.~(94) but misses the square of the factor $(\Lambda_c/\Lambda)$.}
\begin{align}
 |\vec{a}_\theta| \gg a_0 \left(\frac{\Lambda_c}{\bar{\alpha} \Lambda}\right)^2 \,.
\end{align}

In two-field {\sc SFDM}, analogous constructions still work.
For example, consider
\begin{align}
 \label{eq:Lorig+}
 \mathcal{L}_{\rm{orig}+} = \frac12 \bar{K}_+ + \frac{\Lambda^4}{6 (\Lambda_c^2 + \rho_+^2)^6} \bar{K}_+^3 - \lambda \, \theta_+ \, \rho_b \,,
\end{align}
where $\rho_+$ is a new field with
\begin{subequations}
 \label{eq:Kbar}
\begin{align}
    \bar{K}_+ &= K_{\rho_+} + \rho_+^2 ( K_+ + K_- -m^2) \,, \\
   K_{\rho_+} &= \nabla^\alpha \rho_+ \nabla_\alpha \rho_+ \,.
\end{align}
\end{subequations}
This is the same as $ \mathcal{L}_{\rm{orig}} $ from Eq.~\eqref{eq:Lorigstandard}, but with $ \rho $ replaced by $ \rho_+ $, $ K_\theta $ replaced by $ K_+ + K_- $, and $ \theta $ replaced by $ \theta_+ $ in the baryon coupling.
Then, similar as in standard {\sc SFDM}, we recover the $ f(K_+ + K_- - m^2) $ part of the Lagrangian $ \mathcal{L}_{\rm{imp}} $ of two-field {\sc SFDM} for the solution $ \rho_+^2 = \Lambda \sqrt{|K_+ + K_- - m^2|} $.
This solution is valid if we can neglect derivatives of $\rho_+$ and if $\rho_+^2 \gg \Lambda_c^2$.
As in standard {\sc SFDM}, the condition $\rho_+^2 \gg \Lambda_c^2$ translates to a minimum acceleration.
In the {\sc MOND} limit $ (\vec{\nabla} \theta_+)^2 \gg 2 m \hat{\mu} $,
\begin{align}
 |\vec{a}_+| \gg a_0 \left(\frac{\Lambda_c}{\bar{\alpha} \Lambda}\right)^2 \,.
\end{align}
A possible alternative to the definition of $\bar{K}_+$ from Eq.~\eqref{eq:Kbar} is
\begin{align}
 \label{eq:Kbaralt}
 \bar{K}_+ &= K_{\rho_+} + \rho_+^2 (K_+ - m^2) + (K_{\rho_-} + \rho_-^2 K_-) \cdot \frac{\rho_+^2}{\rho_-^2 + \Lambda_{-}^2} \,,
\end{align}
with constant $\Lambda_{-}$.
For $\rho_-^2 \gg \Lambda_{-}^2$ and as long as derivatives of $\rho_-$ can be neglected, this gives the same phenomenology as Eq.~\eqref{eq:Kbar}.
However, in contrast to our earlier definition, it is now straightforward to rewrite $\bar{K}_+$ in terms of $\phi_- \propto \rho_- e^{-i \theta_-}$, since $\nabla_\alpha \phi_-^* \nabla^\alpha \phi_- \propto K_{\rho_-} + \rho_-^2 K_- $.

\section{Calculation of sound speeds}
\label{sec:soundspeed:calculation}

In Eq.~\eqref{eq:Limppert}, we calculated second-order perturbations to the Lagrangian $ \mathcal{L}_{\rm{imp}} $, but without the contributions from $ \mathcal{L}_- $.
For the stability analysis in Sec.~\ref{sec:model}, this was sufficient.
To calculate the sound speeds, we also need the contributions from $ \mathcal{L}_- $,
\begin{align}
\begin{split}
 \mathcal{L}_{-\rm{pert}} = \frac12 g^{\alpha \beta} \partial_\alpha \delta_\rho \partial_\beta \delta_\rho + \frac12 g^{\alpha \beta} \rho_0^2 \partial_\alpha \delta_- \partial_\beta \delta_- - (\mu_0^2 g^{00}-m^2) \delta_\rho^2 + 2 g^{00} \mu_0 \rho_0 \delta_\rho \dot{\delta}_- \,,
\end{split}
\end{align}
where $ \delta_\rho $ is the perturbation of $ \rho_- = \rho_0 + \delta_\rho $ with $ \rho_0^2 = (g^{00} \mu_0^2 - m^2)/\lambda_4 $.
Note that the $ \delta_\rho \dot{\delta}_- $ term does not enter the Hamiltonian and thus does not affect the stability analysis.
This is why we neglected this contribution in the stability analysis in Sec.~\ref{sec:model}.

We now restrict our attention to perturbations with wavelengths that are short compared to the background field gradients.
That is, we approximate terms like $ \partial ( X_0 \delta ) $ as $ X_0 (\partial \delta) $, where $ \delta $ is a perturbation and $ X_0 $ is a function of background fields only.
Roughly, on galactic scales, this is valid for perturbations with wave vector $ |\vec{k}| \gg (1/\rm{kpc}) $.
Then, adding up all contributions gives:
\begin{align}
\label{eq:Limppertfull}
\begin{split}
 \mathcal{L}_{\rm{imp},\rm{pert}}
 = &+\frac12 \left(\dot{\delta}_\rho^2 - (\vec{\nabla} \delta_\rho)^2 \right) - (\mu_0^2 g^{00} -m^2) \delta_\rho^2 \\
   &+\left(f_0' \dot{\delta}_+^2 - (f_0' - 2 f_0'' \gamma^2 |\vec{\nabla} \theta_+|^2)  (\vec{\nabla} \delta_+)^2 \right) \\
   &+\left((f_0' + 2 f_0'' \mu_0^2 + A) \dot{\delta}_-^2 - (f_0' + A) (\vec{\nabla} \delta_-)^2 \right) \\
   &+\left(- 4 f_0'' \mu_0 |\vec{\nabla} \theta^0_+| \gamma\right) |\vec{\nabla} \delta_+| \dot{\delta}_- \\
   &+2\mu_0 \rho_0 \delta_\rho \dot{\delta}_- \,,
\end{split}
\end{align}
where $ A = \rho_0^2/2 $, see Sec.~\ref{sec:model}, and $ \gamma $ is the cosine of the angle between $ \vec{\nabla} \theta^0_+ $ and $ \vec{\nabla} \delta_+ $,
\begin{align}
 \vec{\nabla} \theta^0_+ \cdot \vec{\nabla} \delta_+ \equiv \gamma |\vec{\nabla} \theta^0_+| |\vec{\nabla} \delta_+| \,.
\end{align}
Note that $ \gamma $ is not a constant and depends on both $ \delta_+ $ and $ \theta^0_+ $.
To find the sound speeds, we consider ansätze $ \delta_j \to \delta_j e^{-ikx} + \delta_j^* e^{ikx} $ for $ j = +,-,\rho$ with $ e^{ik x } = e^{i(\omega t - \vec{k} \vec{x})} $.

\subsection{Without mixing of $ \delta_+ $ and $ \delta_- $}

We first neglect the mixing of $ \delta_+ $ and $ \delta_- $, i.e. we neglect the fourth line in Eq.~\eqref{eq:Limppertfull}.
Then, the $ \delta_+ $ equation of motion directly gives
\begin{align}
 c_{s,+}^2 \approx \frac{f_0' - 2 f_0'' \gamma^2 |\vec{\nabla} \theta^0_+|^2}{f_0'} \approx 1 + \gamma^2 \geq 1 \,,
\end{align}
where we used $ - 2 f_0'' |\vec{\nabla} \theta^0_+|^2 \approx f_0' $, which follows from Eq.~\eqref{eq:f0imp} in the {\sc MOND} limit $ (\vec{\nabla} \theta^0_+)^2 \gg 2 m \hat{\mu}_0 $.
The other gapless mode is the standard superfluid phonon mode which is a mix of $ \delta_- $ and $ \delta_\rho $.
Their two equations of motion are:
\begin{align}
 0 = &\left(\omega^2 - \vec{k}^2 - 2 (\mu^2 g^{00} - m^2)\right) (\delta_\rho e^{i k x} + \delta_\rho^* e^{-ikx}) \\
     & + 2 i \omega \mu_0 \rho_0 (\delta_- e^{ikx} - \delta_-^* e^{-ikx})  \nonumber \,, \\
 0 = &\left(2 (f_0' + 2 f_0'' \mu^2 + A) \omega^2 - 2 (f_0' + A) \vec{k}^2 \right) (\delta_- e^{i k x} + \delta_-^* e^{-ikx}) \\
     & - 2 i \omega \mu_0 \rho_0 (\delta_\rho e^{ikx} - \delta_\rho^* e^{-ikx}) \,. \nonumber
\end{align}
The first is the $ \delta_\rho $ equation, the second is the $ \delta_- $ equation.
From the $ \delta_\rho $ equation, we find:
\begin{align}
 \delta_\rho = - \frac{2 i \omega \mu_0 \rho_0}{\omega^2 - \vec{k}^2 -2 (\mu^2 g^{00} - m^2)} \delta_- \,.
\end{align}
Note that this relation is still valid when we later allow for $\delta_+$-$\delta_-$-mixing, since this mixing only modifies the $ \delta_- $ equation, but not the $ \delta_\rho $ equation.
Plugging this in the $\delta_-$ equation and dividing by $A = \rho_0^2/2$ gives
\begin{align}
 \label{eq:csnomix}
 0 &= \left(2 \left(\frac{f_0' + 2 f_0'' \mu_0^2}{A} + 1\right) \omega^2 - 2 \left(\frac{f_0'}{A} + 1\right) \vec{k}^2 \right) - \frac{8 \mu_0^2 \omega^2}{\omega^2 - \vec{k}^2 -2 (\mu_0^2 g^{00} - m^2)} \,.
\end{align}
In Sec.~\ref{sec:model}, we saw that typically $ A \gg f_0' $.
Thus, the prefactor of $ \vec{k}^2 $ in this equation reduces to  $ -2 $.
Similarly, we saw in Sec.~\ref{sec:model}, that we need $ A > |2 f_0'' \mu_0^2 | $ for stability.
Thus, the first term in Eq.~\eqref{eq:csnomix} contributes a term $ \mathcal{O}(1) \cdot \omega^2 $.
This is negligible compared to the last term, which is $ 2 m \omega^2 /\hat{\mu}_0 \gg \omega^2 $ for small wavevectors.
Thus, we recover the standard phonon dispersion relation \cite{Schmitt2015}
\begin{align}
 c_{s-}^2 = \frac{\hat{\mu}_0}{m} \,.
\end{align}
Note that the dominant prefactor of $ \omega^2 $ in Eq.~\eqref{eq:csnomix} comes from the $\delta_\rho$-$\delta_-$-mixing.
Thus, this mixing cannot be neglected.

\subsection{With mixing of $ \delta_+ $ and $ \delta_- $}

With the $ \delta_+ $-$ \delta_- $ mixing term, we have for $ \delta_+ $:
\begin{align}
\begin{split}
 0 = &\left( 2 f_0' \omega^2 - 2 (f_0' - 2 f_0'' \gamma^2 |\vec{\nabla} \theta^0_+|^2) \vec{k}^2 \right) (\delta_+ e^{ikx} + \delta_+^* e^{-ikx})  \\
 &+ \left(4 f_0'' \mu_0 |\vec{\nabla} \theta^0_+| \gamma\right) |\vec{k}| \omega (\delta_- e^{ikx} + \delta_-^* e^{-ikx}) \,.
\end{split}
\end{align}
And for $ \delta_- $
\begin{align}
\begin{split}
 0 = & (\delta_- e^{ikx} + \delta_-^* e^{-ikx}) \begin{aligned}[t] \left[\left( 2(f_0' + 2 f_0'' \mu^2 + A) \omega^2 - 2(f_0' + A) \vec{k}^2 \right) \right. \\
                                                                   \left. - \frac{(2 \mu_0 \rho_0 \omega)^2}{\omega^2 - \vec{k}^2 -2 (\mu_0^2 g^{00} - m^2)}\right] \end{aligned} \\
 &+ \left(4 f_0'' \mu_0 |\vec{\nabla} \theta^0_+| \gamma\right) |\vec{k}| \omega (\delta_+ e^{ikx} + \delta_+^* e^{-ikx}) \,,
\end{split}
\end{align}
where we already plugged in the solution for $ \delta_\rho $ from the previous section.
As discussed there, this solution is still valid here.
Solving the $\delta_+$ equation for $ \delta_+ $ gives
\begin{align}
 \label{eq:deltapmmix}
 \delta_+ = - \delta_- \frac{4 f_0'' \mu_0 |\vec{\nabla} \theta^0_+| \gamma |\vec{k}| \omega}{2 f_0' \omega^2 - 2(f_0' - 2f_0'' \gamma^2 |\vec{\nabla} \theta^0_+|^2)\vec{k}^2} \,.
\end{align}
Plugging this result into the $\delta_-$ equation and dividing by $ A$ yields
\begin{align}
\begin{split}
 0 = &\left( 2\left(\frac{f_0' + 2 f_0'' \mu_0^2}{A} + 1\right) \omega^2 - 2\left(\frac{f_0'}{A} + 1\right) \vec{k}^2 \right) - \frac{8 \mu_0^2 \omega^2}{\omega^2 - \vec{k}^2 -2 (\mu_0^2 g^{00} - m^2)} \\
     &- \frac{2 f_0'' \mu_0^2}{A} \frac{4 \cdot 2 f_0'' |\vec{\nabla} \theta^0_+|^2 \gamma^2 |\vec{k}|^2 \omega^2}{2 f_0' \omega^2 - 2(f_0' - 2 f_0'' \gamma^2 |\vec{\nabla} \theta^0_+|^2) \vec{k}^2} \,.
\end{split}
\end{align}
Numerically, $ 2 f_0'' |\vec{\nabla} \theta^0_+|^2 \approx -f_0' $.
So we can write:
\begin{align}
\begin{split}
 \label{eq:dispersion}
 0 = &\left( 2\left(\frac{f_0' + 2 f_0'' \mu_0^2}{A} + 1\right) \omega^2 - 2\left(\frac{f_0'}{A} + 1\right) \vec{k}^2 \right) - \frac{8 \mu_0^2 \omega^2}{\omega^2 - \vec{k}^2 -2 (\mu_0^2 g^{00} - m^2)} \\
     &-\left(- \frac{2 f_0'' \mu_0^2}{A}\right) \frac{2 \cdot \gamma^2 |\vec{k}|^2 \omega^2}{ \omega^2 - (1 + \gamma^2) \vec{k}^2} \,.
\end{split}
\end{align}
From this equation, we can now find the sounds speeds for the two gapless modes.

Consider first the ``$ \delta_- $-and-$\delta_\rho$'' mode with $ \omega = c_{s-} |\vec{k}| $ with $ c_{s-} \ll 1 $.
For $ k \to 0 $, Eq.~\eqref{eq:dispersion} becomes
\begin{align}
\begin{split}
 0 = \vec{k}^2 &\left[ \left( 2\left(\frac{f_0' + 2 f_0'' \mu_0^2}{A} + 1\right) c_{s-}^2 - 2\left(\frac{f_0'}{A} + 1\right) \right) + \frac{4 \mu_0^2 c_{s-}^2}{(\mu_0^2 g^{00} - m^2)} \right. \\
 &\left. -\left(- \frac{2 f_0'' \mu^2}{A}\right) \frac{2 \cdot \gamma^2 }{ 1 - \frac{(1 + \gamma^2)}{c_{s-}^2}} \right] \,.
\end{split}
\end{align}
If we expand the last term in $ c_{s-}^2 $, we find
\begin{align}
 c_{s-}^2 \left( \frac{2 \mu_0^2}{\mu_0^2 g^{00} - m^2} + \left(\frac{f_0'}{A}+1 + \frac{2f_0'' \mu_0^2}{A}\right) + \left(-\frac{2 f_0'' \mu_0^2}{A} \frac{\gamma^2}{1+\gamma^2} \right) \right) = \left(1+ \frac{f_0'}{A}\right) \,.
\end{align}
For a non-relativistic background, we have $ 2 \mu_0^2 / (\mu_0^2 g^{00} - m^2) \approx m/\hat{\mu}_0 $ and we can expand in $\hat{\mu}_0/m \ll 1$:
\begin{align}
\begin{split}
 c_{s-}^2 &= \frac{1+\frac{f_0'}{A}}{ \frac{2 \mu_0^2}{\mu_0^2 g^{00} - m^2} + \left(\frac{f_0'}{A}+1 + \frac{2f_0'' \mu_0^2}{A} \frac{1}{1+\gamma^2}\right) } \\
       &= \frac{\hat{\mu}_0}{m} \left(1+ \frac{f_0'}{A}\right) \left[1 - \frac{\hat{\mu}_0}{m} \left(1 + \frac{f_0'}{A}  + \frac{2 f_0'' \mu_0^2}{A} \frac{1}{1+\gamma^2} \right)  + \dots \right] \,.
\end{split}
\end{align}
Since $A \gg f_0'$ and $A > |2 f_0'' \mu_0^2| $ on galactic scales, this gives only small corrections to the value of $c_{s-}$ without mixing of $ \delta_+ $ and $ \delta_- $, namely $ c_{s-}^2 = \hat{\mu}_0/m $.

For the ``$ \delta_+ $'' mode, which had $ \omega^2 = (1 + \gamma^2) \vec{k}^2 $ without $\delta_+$-$\delta_-$-mixing, we consider again non-relativistic perturbations, $ |\vec{k}| \ll m $, and a non-relativistic background, $ |\mu_0^2 g^{00} - m^2| \ll m^2 $.
Then, $ 4 \mu_0^2/(\omega^2 - \vec{k}^2 - 2(\mu_0^2 g^{00} - m^2)) \gg 1 $ so that we can neglect the first two terms in \eqref{eq:dispersion},
\begin{align}
 0 = - \frac{4 \mu_0^2}{\omega^2 - \vec{k}^2 - 2(\mu_0^2 g^{00} - m^2)} \omega^2 - \left(-\frac{2 f_0'' \mu_0^2}{A}\right) \frac{\gamma^2 |\vec{k}|^2 \omega^2}{\omega^2 - (1 + \gamma^2) \vec{k}^2} \,.
\end{align}
With the Ansatz $ \omega^2 = c_{s,+}^2 \vec{k}^2 $, this yields
\begin{align}
 \label{eq:cs+withmixing}
 c_{s,+}^2 = 1 + \gamma^2\left[1 + \eta \frac{2 (\mu_0^2g^{00} - m^2) - \gamma^2 \vec{k}^2}{4 \mu_0^2+ \gamma^2 \vec{k}^2 \eta}\right] \geq 1 \,,
\end{align}
where we defined
\begin{align}
 \eta \equiv - \frac{2 f_0'' \mu_0^2}{A} \,.
\end{align}
As mentioned above, for stability on galactic scales we need $ A > |2 f_0'' \mu_0^2| $.
Thus, we typically have $ \eta \in (0, 1) $ on galactic scales.
Then, for $ |\vec{k}| \ll m $ and a non-relativistic background, Eq.~\eqref{eq:cs+withmixing} gives only small corrections to the result without mixing of $ \delta_+ $ and $ \delta_- $, namely $ c_{s+}^2 = 1 + \gamma^2 $.

\end{appendices}

\printbibliography[heading=bibintoc]

\end{document}